\documentclass[sigconf]{acmart}
\usepackage{multirow}
\usepackage{graphicx}
\usepackage{array} 
\usepackage{cleveref}
\usepackage{listings}
\usepackage{enumitem}
\AtBeginDocument{%
  }

\setcopyright{acmlicensed}

\copyrightyear{2026}
\acmYear{2026}
\setcopyright{cc}
\setcctype{by}
\acmConference[CHI '26]{Proceedings of the 2026 CHI Conference on Human Factors in Computing Systems}{April 13--17, 2026}{Barcelona, Spain}
\acmBooktitle{Proceedings of the 2026 CHI Conference on Human Factors in Computing Systems (CHI '26), April 13--17, 2026, Barcelona, Spain}
\acmPrice{}
\acmDOI{10.1145/3772318.3790304}
\acmISBN{979-8-4007-2278-3/2026/04}

\acmSubmissionID{1899}

\begin{document}


\title{\ouyangre{\textit{CaseMaster}: Designing and Evaluating a Probe for Oral Case Presentation Training with LLM Assistance}}

\author{Yang Ouyang}
\orcid{0009-0000-5841-7659}
\affiliation{%
  \institution{School of Information Science and Technology \\ ShanghaiTech University}
  \city{Shanghai}
  \country{China}
}
\email{ouyy@shanghaitech.edu.cn}

\author{Yuansong Xu}
\orcid{0009-0005-1630-6279}
\affiliation{%
  \institution{School of Information Science and Technology \\ ShanghaiTech University}
  \city{Shanghai}
  \country{China}
}
\email{xuys2023@shanghaitech.edu.cn}

\author{Chang Jiang}
\orcid{0000-0002-7468-3372}
\affiliation{
  \institution{Department of Clinical Medicine \\ Shanghai Clinical Research and Trial Center}
  \city{Shanghai}
  \country{China}
}
\email{cjiang_fdu@yeah.net}

\author{Yifan Jin}
\orcid{0009-0007-3839-7612}
\affiliation{%
  \institution{School of Information Science and Technology \\ ShanghaiTech University}
  \city{Shanghai}
  \country{China}
}
\email{jinyf2024@shanghaitech.edu.cn}

\author{Haoran Jiang}
\orcid{0009-0009-5717-4208}
\affiliation{%
  \institution{School of Information Science and Technology \\ ShanghaiTech University}
  \city{Shanghai}
  \country{China}
}
\email{jianghr2023@shanghaitech.edu.cn}

\author{Quan Li}
\authornote{Corresponding Author.}
\orcid{0000-0003-2249-0728}
\affiliation{%
  \institution{School of Information Science and Technology \\ ShanghaiTech University}
  \city{Shanghai}
  \country{China}
}
\email{liquan@shanghaitech.edu.cn}

\renewcommand{\shortauthors}{Ouyang et al.}
\newcommand{\ouyangrevision}{\textcolor{black}}

\newcommand{\ouyangre}{\textcolor{black}}
\newcommand{\ouyangrese}{\textcolor{teal}}

\begin{abstract}
Preparing an oral case presentation (OCP) is a crucial skill for medical students, requiring clear communication of patient information, clinical findings, and treatment plans. However, inconsistent student participation and limited guidance can make this task challenging. While Large Language Models (LLMs) can provide structured content to streamline the process, their role in facilitating skill development and supporting medical education integration remains underexplored. To address this, we conducted a formative study with six medical educators and developed \textit{CaseMaster}, an interactive probe that leverages LLM-generated content tailored to medical education to help users enhance their OCP skills. The controlled study suggests \textit{CaseMaster} has the potential to both improve presentation quality and reduce workload compared to traditional methods, an implication reinforced by expert feedback. We propose guidelines for educators to develop adaptive, user-centered training methods using LLMs, while considering the implications of integrating advanced technologies into medical education.
\end{abstract}




\maketitle
\section{INTRODUCTION}
\par As a standardized format for exchanging clinical information, the oral case presentation (OCP) is essential for communicating patient details, exam findings, diagnostic rationale, and treatment strategies among physicians~\cite{green2009expectations,packer2019presenting}. Its importance as an educational goal for learners has been recognized by the Clerkship Directors in Internal Medicine (CDIM)~\cite{goroll1998core} and many other medical educators~\cite{accreditation2007outcome,bass1997national}. A well-executed case presentation enhances the clarity and efficiency of information transfer between physicians, ensuring that critical patient details are accurately communicated.

\par To achieve proficiency in OCP, medical educators emphasize the necessity of robust training methods, employing techniques such as case discussions~\cite{reifenrath2022integrated}, case writing~\cite{rison2013guide}, and role-playing~\cite{nestel2007role} individually or in combination to maximize learning outcomes and practical skills. In practice, students participate in case discussions by analyzing clinical cases with a facilitator, draft and review detailed reports in case writing before oral presentations, and simulate clinical scenarios through role-playing exercises, acting as either doctors or patients. These approaches collectively enhance students' OCP abilities effectively~\cite{mclean2016case}.

\par However, several challenges can hinder the effectiveness of the above-mentioned methods. First, while these approaches are beneficial for developing critical thinking and communication skills, they can sometimes be dominated by more vocal students, leaving quieter individuals less engaged~\cite{bradley2012contemporary,Verma2021Interactive,Melvin2016The}. This imbalance can create a learning environment where not all students benefit equally. Although techniques like Round-Robin Participation~\cite{delina2021application} and the Six Thinking Hats method~\cite{Yang2014Application} can help increase student engagement, some students still lack sufficient training. Additionally, while numerous studies have developed training systems that provide automated feedback on verbal communication~\cite{damian2015augmenting,tanveer2015rhema,trinh2017robocop,wang2020voicecoach}, these systems primarily focus on public speaking and often require users to specify parameters such as spoken emphasis or voice modulation. Although these aspects are important, they do not fully address the specific challenges of OCPs. Second, educators expect students to apply clinical reasoning skills to select relevant details from a patient's history, physical examination, and ancillary studies while employing rhetorical skills to enhance the organization and clarity of the presentations. However, students often struggle due to unclear expectations from instructors regarding OCPs. Without clearly defined initial goals and quantitative feedback, students find it challenging to gauge their performance and make necessary improvements. This lack of direction and measurable feedback can impede their learning process, leading to uncertainty and reduced confidence in their presentation skills. Furthermore, implementing these methods effectively require significant time and resources~\cite{Green2005Developing}. In a curriculum already packed with theoretical and practical knowledge, finding the balance to help students develop their oral case presentation skills can be challenging~\cite{carter2008study}. Instructors may struggle to provide adequate hands-on training, which is not always feasible~\cite{Heiman2012E-learning}.

\par Numerous studies have explored the use of large language models (LLMs) in educational settings due to their ability to offer personalized, efficient, and engaging learning and teaching experiences~\cite{mogavi2023exploring,abd2023large,safranek2023role,benitez2024harnessing,lucas2024systematic}. These advancements hold significant potential for addressing challenges in formulating instructional strategies by providing in-situ-generated content that adapts to educators' insights~\cite{mogavi2024chatgpt,mollick2023using}, thereby facilitating the development of OCP skills. However, without structured integration, the use of LLMs can lead to fragmented training experiences, as the output may lack systematic organization or direct relevance to educational objectives. While some studies~\cite{yan2024knownet,pan2024unifying} have integrated LLMs with Knowledge Graphs (KG) to enhance accuracy and provide more structured exploration, these approaches may face challenges when dealing with complex, implicit human knowledge, particularly in less structured or highly context-dependent medical scenarios~\cite{ji2021survey,zhao2022simulate}. Furthermore, there remains a gap in understanding the design requirements for such skills training, how LLMs can effectively meet these needs, and how medical practitioners would perceive and collaborate with these technologies.

\par \ouyangrevision{In response, we investigated the effectiveness of structured training frameworks in equipping medical practitioners to learn OCPs and identifying any additional support they may require. We collaborated with six educators to evaluate educational interventions aimed at improving students' OCP skills.} 
As part of this collaborative effort, we developed a system probe named \textit{CaseMaster} to assist medical students in training their OCP skills. The probe guides users through preparation and reflection stages, enabling them to explore patient cases, prepare using the SOAP format with LLM-generated content, and reflect by comparing their solutions with reference answers, integrating feedback to enhance their skills and confidence in a structured learning experience. We then evaluated the effectiveness and user experience of the probe through two user studies. The first study, involving $12$ medical students using our probe and a baseline tool, suggested positive impacts on confidence and presentation skills. The second study, with five experienced educators, provided valuable feedback, helping refine our probe for broader application in medical education.

\par This study presents three key contributions. \ouyangre{First, we conduct an in-depth analysis of how to effectively design and implement training for medical students to master OCP, a pivotal skill in medical communication. Based on these findings, we develop a probe tailored to OCP skills, incorporating instructional strategies to support the development of students' training objectives. 
Second, user study results suggest that the tool, through its integration with LLMs, has the potential to improve the clarity, accuracy, and overall quality of medical students' oral case presentations. Finally, we propose a set of design and implementation guidelines for developing effective training tools with LLMs. These guidelines emphasize evidence-driven methodologies, dynamic engagement, and integrative frameworks to fully harness the potential of advanced technologies in medical education.}

\section{BACKGROUND and RELATED WORK}
\subsection{Oral Case Presentation in Medicine}
\par The oral case presentation (OCP) is a crucial communication skill in medicine~\cite{green2009expectations,packer2019presenting}. Studies have examined educator expectations for OCP within the field~\cite{green2009expectations,Murphy2018ActingLA}. For instance, Green et al.~\cite{green2009expectations} conducted a significant survey of $110$ members of the CDIM to understand common expectations for OCPs among clinical clerks. Other research has explored various methods to enhance OCP skills. Kim et al.~\cite{kim2005randomized} conducted a randomized controlled study that demonstrated how using encounter cards, which provided structured and timely feedback, effectively improved OCP presentation skills among medical students. Chan et al.~\cite{Chan2015TheOC} reviewed medical educators' expectations for effective case presentations, highlighted the absence of a comprehensive model for teaching OCPs, and advocated for a performance-based approach to enhance structured feedback and teaching support for medical students. 
Simultaneously, parallel research efforts focus on developing frameworks and curricula to improve OCP skills among medical students~\cite{daniel2015teaching,heiman2014improving}. For example, Williams et al.~\cite{Williams2016DevelopingOC} constructed a curriculum designed to improve medical students' oral case presentation skills through instruction, self-evaluation, and peer evaluation, with the goal of improving quality while reducing the time burden on faculty. Wawrykow et al.~\cite{Wawrykow2020MP16OC} developed and assessed a competency-based tool and curriculum specifically for OCPs in emergency medicine. \ouyangrevision{Ouyang et al.~\cite{ouyang2025casemaster} developed a system to support OCP training with LLM guidance, but its impact on students' effectiveness and overall user experience remains unexamined.}

\par These studies highlight the ongoing efforts to standardize and improve educational strategies for training OCP skills. The alignment in educator expectations suggests that a unified approach to training and evaluating OCPs could be highly beneficial. 

\subsection{Intelligent Instructional Tools}
\par To facilitate knowledge acquisition and skill development, HCI researchers have developed a range of intelligent instructional tools. For instance, Wambsganss et al.~\cite{10.1145/3411764.3445781} introduced \textit{ArgueTutor}, a system that dynamically adapts to users' needs by providing real-time feedback to improve their argumentative writing, such as pinpointing areas for improvement in their drafts. Similarly, Ruan et al.~\cite{10.1145/3290605.3300587} developed \textit{QuizBot}, an AI-powered chatbot designed to help students master factual knowledge in subjects like science, safety, and English vocabulary. This tool actively engages learners by posing questions and offering corrective feedback based on their responses. Additionally, Zhang et al.~\cite{Zhang2023VISARAH} designed \textit{VISAR}, enabling users to explore, experiment with, and validate their writing plans using automatic draft prototyping. These intelligent instructional systems provide adaptive support to enhance the learning experience. Building on these concepts and focusing on medical education, we offer first-hand insights into the design, effectiveness, and user experience of an adaptive training probe specifically for OCPs.

\subsection{Training Systems for Enhancing Presentation Skills}
\par Given the strong similarities between OCP skills and general presentation training-both emphasizing clear communication, effective delivery, and audience engagement-we reviewed existing studies in training systems to inform our approach. 
\ouyangre{Various systems in the HCI field provide automated feedback on presentation quality by analyzing vocal features such as pitch, speech rate, and loudness~\cite{Kurihara2007PresentationSA,Ochoa2020ControlledEO}. For example, Zhang et al.~\cite{10.1145/3313831.3376322} introduced \textit{Withyou}, an adaptive speech shadowing tool designed to aid foreign language learning. However, many existing systems rely on fixed thresholds for assessment, which often lack the contextual nuance needed for targeted improvements~\cite{rubin2015capture}. In addition, the feedback strategies used can significantly influence training outcomes. While real-time feedback offers immediate insights that help users correct mistakes quickly~\cite{damian2015augmenting,schneider2015presentation,schneider2017presentation,tanveer2015rhema}, research also suggests that intermittent feedback plays a crucial role in enhancing long-term retention and fostering sustained learning and skill mastery~\cite{schimdt2019motor,lowe2017impact}. Interactive offline systems, such as \textit{EmoCo}~\cite{zeng2019emoco} and \textit{VoiceCoach}~\cite{wang2020voicecoach}, further support self-reflection and iterative improvement by integrating analytics with user interaction.} 
\par \ouyangrevision{
Similarly, medical training research has investigated approaches for structured peer interaction. Handoff and team systems, for instance, use simulation and guided practice to facilitate collaborative activities~\cite{king2008teamstepps,berkenstadt2008improving,starmer2014changes}. However, these tools primarily target protocolized, safety-critical workflows rather than the clinical-reasoning–driven, physician-to-physician presentations characteristic of OCPs~\cite{chan2015oral}.}


\par Building on these insights, our approach emphasizes effective content delivery in OCP skill training, prioritizing practical, problem-oriented presentations. Unlike traditional methods that focus on vocal metrics like tone and speed to engage the audience, we place greater emphasis on content structure and coherence, ensuring the logical organization and clarity of key points for more effective communication and understanding of essential information.

\subsection{Large Language Models in Education}

\par \ouyangre{Recent advancements have shifted the focus of educational applications towards LLMs, which offer personalized, efficient, and engaging teaching and learning experiences~\cite{mogavi2023exploring}. LLMs can generate tailored self-study materials, such as flashcards and course-specific questions, to meet individual learning needs~\cite{chen2024retassist,rahman2023chatgpt,bhat2022towards}. For example, Divito et al.~\cite{divito2024tools} used ChatGPT to support medical learners in problem-based learning by clarifying complex queries~\cite{kung2023performance}. Huber et al.~\cite{Huber2024LeveragingTP} introduced game-based methods to enhance skill development while mitigating over-reliance. Moreover, LLMs facilitate critical thinking through interactive, open-ended learning experiences~\cite{denny2024generative,abdelghani2022gpt}.}


\par \ouyangre{On the other hand, LLMs offer educators powerful tools to streamline assessments and deliver personalized feedback~\cite{bernius2022machine,xia2022persua}.  For instance, Mitra et al.~\cite{mitra2024retllm} developed \textit{RetLLM-E} to assist educators in compiling and addressing frequently asked student questions by synthesizing responses from prior discussions and course materials, enabling LLMs to generate high-quality, precise answers. Macneil et al.\cite{Macneil2022TheIO} emphasized the role of LLMs in fostering inclusivity and awareness in computer science education. Additionally, LLMs assist in creating diverse teaching resources, such as instructional notes~\cite{bonner2023large}, explanations~\cite{mollick2023using}, exercises~\cite{shen2021generate}, and FAQs~\cite{mitra2024retllm}. However, there is limited research on how medical practitioners can utilize LLMs to optimize workflows in medical education. Traditional education methods for OCP training rely heavily on practical experience, such as case discussions~\cite{reifenrath2022integrated} and role-playing~\cite{nestel2007role}. The integration of LLMs offers promising opportunities to streamline training and workflow processes; however, it also raises concerns around their reliability and adaptability in educational contexts. This work aims to investigate these challenges and outline comprehensive design guidelines for developing effective LLM-based training tools in medical education. }

\section{FORMATIVE STUDY}
\par \ouyangre{To design the probe and investigate how LLMs could support the process, we conducted a formative study with six medical educators.} The participants, with an average age of 29.83 (SD = 5.15), included three females and three males, identified as P1 through P6. Half of the participants (P2, P5, P6) are novice educators with less than three years of teaching experience, while the others (P1, P3, P4) have over five years of experience guiding junior students. All participants had prior experience in medical education, ensuring their expertise was directly aligned with our goal of enhancing OCP skills.

\subsection{Procedure}
\label{sec:proce}
\par We divided the procedure into two phases to deepen our understanding of how medical professionals prepare and deliver OCPs, with a focus on user experience and practical needs.

\par In the first phase, participants engaged in semi-structured interviews lasting approximately $45$ minutes. These discussions aimed to uncover the methods and practices employed in the routine preparation of OCPs. Questions explored various aspects, such as whether participant follow a specific templates when organizing case information (e.g., ``\textit{Do you have a specific structure you follow when preparing case information?}''), the types of resources they consult (e.g., ``\textit{Which materials do you find most useful---textbooks, specialized monographs, or online databases?}''), and how they integrate educational theories into their presentations (e.g., ``\textit{How do you apply theories like \textit{SOAP}~\cite{seo2016pilot} in your presentations?}''). The challenges encountered during the preparation process were also examined to identify obstacles that may hinder effective OCP (e.g., ``\textit{What are the most significant challenges you face in preparing for an oral presentation?}''). \ouyangre{Additionally, we explored their perspectives on utilizing LLMs, such as ChatGPT, in designing comprehensive OCPs and how LLMs could assist in structuring and supporting the design of training sessions.}

\begin{figure*}[h]
    \centering
    \includegraphics[width=\textwidth]{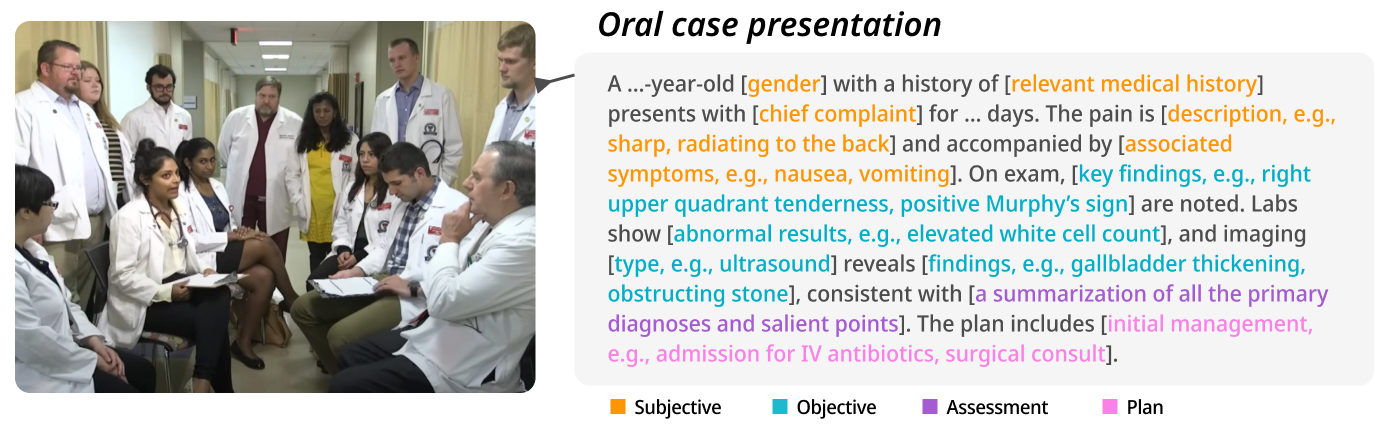}
    \caption{\ouyangre{Example of an oral case presentation: On the left, a group of healthcare professionals, including doctors and medical students, are gathering in a clinical hallway for a case presentation, with the presenter leading the discussion while the others listen attentively. On the right, the case presentation is structured into four key sections: Orange (Subjective) for history and symptoms, Blue (Objective) for findings, Purple (Assessment) for diagnosis, and Pink (Plan) for treatment.}}
    \label{fig:case}
\end{figure*}

\par \ouyangrevision{Following the interviews, participants engaged in a 35-minute co-design session to refine and evaluate preliminary concepts for OCPs, structured into three stages: a 10-minute exploration, a 15-minute manipulation, and a 10-minute reflection (see \cref{fig:codesign}). During exploration, participants discussed their current practices in preparing and delivering OCPs, sharing challenges, priorities, and strategies, which allowed the researchers to surface implicit requirements and identify opportunities for support in both preparation and presentation. In the manipulation stage, participants worked directly with rough sketches and mockups in Miro\footnote{\url{https://miro.com/app/dashboard/}}
, reorganizing components, annotating key points, and proposing additional functionalities. This hands-on engagement enabled them to externalize their reasoning about workflow preferences, information structuring, and potential support mechanisms. Finally, in the reflection stage, participants articulated the rationale behind their modifications and discussed how different features might influence learner preparation and performance. The sketches were iteratively updated in real time based on participants’ inputs, enabling immediate feedback and refinement. This integrated cycle of discussion, manipulation, and reflection clarified educators’ expectations and informed practical design directions for the probe.}

\subsection{Data Analysis}
\ouyangre{The entire process was video-recorded and transcribed for thorough analysis. \ouyangrevision{We began by employing affinity diagrams~\cite{harboe2015real,lucero2015using} to systematically categorize and organize the data collected in the first phase of the study. The first author led this process, and all authors met across three meetings to review the emerging themes and reach consensus on the key findings.}
We asked participants in later sessions for their views on findings from earlier ones, enriching our understanding and helping to validate the results. Cross-referencing with the Miro board further contributed a rigorous, iterative analysis, offering deeper insights into participant profiles and the challenges they encounter in practice. This informed the interpretation of their responses.}
\par \ouyangre{After completing the analysis, participants were involved to review the results and assess how well they aligned with the original intentions. They were asked to annotate and provide feedback on aspects that either reflected or contradicted their experiences. This feedback allowed us to refine our analysis, ensuring it more accurately represented the participants' perspectives. Finally, two external medical educators reviewed our findings to support alignment with current practices and uncover blind spots.}

\subsection{Findings}

\par As illustrated in \cref{fig:case}, this represents a typical example of an oral case presentation. All participants emphasized the importance of comprehensive preparation for OCPs. The process typically begins with retrieving guidelines and summarizing key points. The time invested in preparation varies depending on the complexity of the case and the participant's experience, with seasoned educators generally spending less time than novices. During preparation, they refer to a range of resources, including textbooks, specialized manuals, and consensus guidelines, while literature is consulted less frequently. To ensure the case information is well-organized, participants consistently use specific templates or structures. All participants indicated that they primarily prepare for OCPs using notes, typically in Word documents or PowerPoint slides. P2 and P4 mentioned that they occasionally take screenshot from textbook and annotate the text to remind themselves of additional information, such as explanations or supplemental examples. They unanimously agreed that having a document outlining the main flow and content of the presentation was beneficial, as it allowed them to ``\textit{easily conduct the presentation based on the documented plan}'' (P6).

\subsubsection{Challenges in Traditional Practice}
\par We gathered the following insights on the challenges of OCPs from the educators' perspectives. They offered valuable feedback on the difficulties they encounter in planning and delivering effective presentations, pinpointing specific areas where novice presenters often struggle.

\par \textbf{\ouyangrevision{C1. Lack of Structured Guidance for Applying the Training Framework.}} Educators often encounter challenges when designing strategies for novice presenters. The most intuitive approach is to have students learn by observing exemplary OCPs or by evaluating and correcting their own presentations afterward. While educators heavily rely on their teaching experience and existing plans, there is a clear need for more structured guidance. They mentioned that the \textit{SOAP} format (Subjective, Objective, Assessment, Plan)~\cite{seo2016pilot} is a clear and understandable framework, yet novices find it difficult to use it flexibly and effectively. ``\textit{They understand the basics of SOAP, but applying it to complex cases is tough for them}'' (P1). Novices often struggle to adapt the framework to the specific context of each case, leading to presentations that lack depth or coherence. ``\textit{Sometimes, they end up sticking too closely to the format and miss out [on] important details}'' (P3).

\par \textbf{\ouyangrevision{C2. Difficulty Identifying and Addressing Common Presentation Mistakes.}} 
Participants frequently observed several recurring errors during OCPs, underscoring the need for better preparation and delivery techniques. A common issue is a lack of focus, where presenters struggle to effectively highlight the main points. Additionally, many presenters fail to refer to their notes during the presentation, leading to omitted details and a disorganized delivery. As P3 noted, ``\textit{Presenters often don't focus on the key points and forget they can refer to their notes}''. Participants also highlighted the challenge of emphasizing key points without overwhelming the audience with too much information. ``\textit{Instead of covering too many [things], it's better to keep the message simple and clear}'' (P5). This requires a deep understanding of the material and the ability to discern what is most important for the audience to retain.

\par \textbf{\ouyangrevision{C3. Limited Access to Timely and Constructive Feedback.}} Traditional feedback on OCPs often comes from critiques by senior doctors or reminders from peers. ``\textit{Students usually get feedback from their supervising doctors}'' (P1). Presenters typically take notes on this feedback to avoid repeating mistakes in future presentations. ``\textit{Students take notes on the feedback and try to use it into their next presentation}'' (P4). Constructive feedback is essential for improving presentation skills. ``\textit{The best way to improve is to keep practicing and refine your skills based on immediate feedback}'' (P3). Additionally, some feedback may be subtle or indirect, and there are occasions when direct feedback is necessary but may not be provided due to time constraints or other issues.


\subsubsection{Challenges on Using LLMs}
\par \ouyangre{To inform our investigation of how LLMs could support oral case presentation training, we surveyed collaborating medical educators about their attitudes toward these technologies.} All participants recognized the benefits of LLMs, like ChatGPT, for enhancing OCP training, provided certain challenges are addressed, including:


\par \textbf{\ouyangrevision{C4. Uncertainty About How to Effectively Integrate LLMs into Training.}} Most participants acknowledged that their understanding of the training principles behind LLMs was limited, often at a very basic level. However, they did not feel that this lack of technical knowledge impeded their ability to design or integrate these technologies into their workflows. Several participants mentioned that they have already started using LLMs to assist with their learning and work. As P1 noted, ``\textit{We should focus on the medical [challenges], not get bogged down in the technical details, because the real challenge is figuring out how we interact with these [tools]}''. P5 emphasized that although their understanding of LLMs is superficial, it has not prevented them from using these tools effectively in their work. They stressed that the primary concern should be on how these tools can enhance their work, seeing LLMs as tools that ``\textit{can be refined to achieve better outcomes, rather than getting too caught up in the technical details}'' (P3).

\par \textbf{\ouyangrevision{C5. High Cognitive Load in Crafting Effective Prompts.}} Participants reflected on their experiences using LLMs like ChatGPT to enhance educational scenarios in medicine. P1 described how engaging in dialogue with the tool helped refine a case presentation, noting that while some responses were surprising, they ultimately improved the quality of his work. However, he acknowledged that ``\textit{finding the right phrasing to get the most useful response}'' was often challenging. Similarly, P2 found ChatGPT effective in identifying flaws in the text, but mentioned that ``\textit{tweaking prompts to reach those insights}'' required considerable effort. While recognizing the potential of LLMs to assist with teaching flow and materials. However, P5 and P6 expressed concerns about the cognitive load, commenting that it can be mentally demanding to think of effective prompts. As P6 observed, ``\textit{Searching for information directly sometimes seems more efficient than querying ChatGPT with specific questions}''.

\par \textbf{\ouyangrevision{C6. Skepticism About the Reliability of LLM Outputs.}} Participants expressed concerns about the reliability of content produced by LLMs, especially regarding the accuracy of the details, which is crucial in medical education. P5 noted, ``\textit{You might get a quick outline, but the details often aren't quite accurate}''. Several participants also questioned the depth of LLMs' knowledge in specialized fields. P3 commented, ``\textit{LLMs can produce some relevant content, but they sometimes miss the crucial nuances [needed] in specialized fields like oncology or neurology, which can lead to misleading or shallow information}''. P6 added, ``\textit{If LLMs don't give me what I need right away, I'd rather not go back and forth trying to refine the results}''.

\subsection{\ouyangrevision{Design Concepts}}


\begin{figure*}[h]
    \centering
    \includegraphics[width=\textwidth]{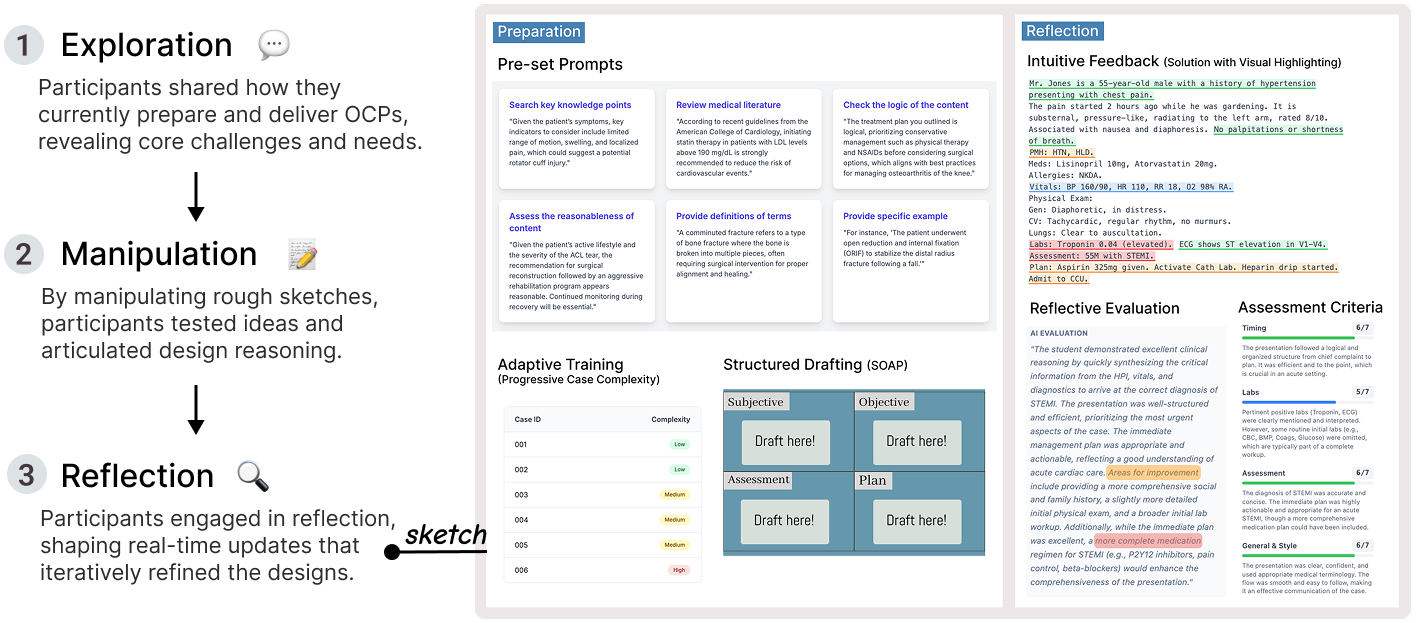}
    \caption{\ouyangrevision{Co-design session focusing on design concepts: participants discussed OCP practices, manipulated sketches to explore ideas, and reflected to refine designs in real time. The design concepts supported both the Preparation and Reflection stages, allowing flexible, low-fidelity exploration and iterative design refinement.}}
    \label{fig:codesign}
\end{figure*}

\par \ouyangrevision{Through our co-design sessions, participants generated a range of design concepts (see \cref{fig:codesign}). We organized these concepts around two recurring stages of the training process: 1) the \textit{Preparation Stage}, where users explore cases and produce initial solution drafts, and 2) the \textit{Reflection Stage}, where users review their reasoning and refine their responses. The concepts include:}

\begin{itemize}

\item \textbf{\ouyangrevision{Preparation Stage (Case Exploration \& Drafting)}}
\begin{itemize}
    \item \ouyangrevision{\textbf{Adaptive Training.} Tiered modules gradually increase case complexity, helping users build confidence and handle progressively challenging scenarios.}
    \item \ouyangrevision{\textbf{Structured Drafting.} Standardized templates (e.g., SOAP) support case construction, ensuring systematic capture of patient information and clear articulation of reasoning.}
    \item \ouyangrevision{\textbf{Pre-set Prompting.} Structured prompts provide guidance on task completion, helping users identify relevant clinical reasoning activities without needing to focus on formatting or query construction.}
\end{itemize}

\item \textbf{\ouyangrevision{{Reflection Stage (Evaluation \& Insight)}}}
\begin{itemize}
    \item \ouyangrevision{\textbf{Visual and Intuitive Feedback.} Real-time visual guidance highlights errors and key points, supporting focused revision and iterative improvement.}
    \item \ouyangrevision{\textbf{Reflective Evaluation.} Users compare drafts with reference responses, receive focused feedback, and examine their decision-making processes.}
    \item \ouyangrevision{\textbf{Transparent Assessment Criteria.} Clear, well-organized rubrics with visual progress indicators clarify performance expectations and help identify strengths as well as areas for further improvement.}
\end{itemize}

\end{itemize}

\subsection{Design Goals}


\par To inform the design of our tool, we synthesized the identified challenges and user-generated concepts to articulate six design goals. These goals represent one plausible design path rather than a definitive or exhaustive set of goals, and informed the design of our probe (see \cref{tab:design_goals}).
\begin{table*}[h]
\centering
\caption{\ouyangrevision{\textbf{Identified challenges and corresponding design goals.}}}
\vspace{-2mm}
\resizebox{\textwidth}{!}{%
\begin{tabular}{cccc} 
\hline
\textbf{OCP Training Stage} & \textbf{Challenges} & \textbf{Design Goals} \\ 
\hline
\multirow{2}{*}{Preparation} & \textbf{C1}. Lack of structured guidance. & \textbf{DG1} \\ 
 & \textbf{C2}. Difficulty identifying and addressing common mistakes. & \textbf{DG2} \\ 
\hline
Reflection & \textbf{C3}. Limited access to timely constructive feedback. & \textbf{DG3} \\ 
\hline
\multirow{3}{*}{Both} & \textbf{C4}. Uncertainty about how to effectively integrate LLMs into training. & \textbf{DG4} \\ 
 & \textbf{C5}. High cognitive load in prompt formulation. & \textbf{DG5} \\ 
 & \textbf{C6}. Skepticism about the reliability of LLM outputs. & \textbf{DG6} \\
\hline
\end{tabular}
}
\label{tab:design_goals}
\end{table*}

\par \ouyangrevision{\textbf{DG1: Support Progressive Skill Development.} To address the need for adaptive training (\textbf{C1}), the system should scaffold engagement with cases of varying complexity, enabling users to gradually build competence.}

\par \ouyangrevision{\textbf{DG2: Promote Systematic Reasoning and Reduce Common Errors.} To mitigate recurring mistakes during OCPs (\textbf{C2}), the system should guide users in structuring case information clearly and reasoning effectively.}

\par \ouyangrevision{\textbf{DG3: Provide Immediate and Constructive Feedback.} Given that delayed or indirect feedback limits performance refinement (\textbf{C3}), the system should deliver actionable guidance that highlights errors and key points, supporting iterative improvement.}

\par \ouyangrevision{\textbf{DG4: Prioritizing Intuitive LLM Design for OCP Training.} Considering users’ limited experience with LLMs (\textbf{C4}) and cognitive demands (\textbf{C5}), the system should provide structured prompts or templates to support reasoning without overwhelming users.}

\par \ouyangrevision{\textbf{DG5: Encourage Reflective Evaluation.} To promote insight and self-assessment (\textbf{C6}), the system should allow users to compare outputs with reference standards and reflect on their reasoning processes.}

\par \ouyangrevision{\textbf{DG6: Facilitate Transparency and User Agency.} The system should provide clear criteria, visual indicators, and flexible interaction to help users understand progress and maintain control over the training process (\textbf{C3–C6}).}


\section{PROBE: CASEMASTER}
\par In this section, we introduce \textit{CaseMaster}, an interactive system probe designed to assist medical students in training their OCP skills using LLMs. \textit{CaseMaster} is implemented as a web application, with the front end developed using Vue 3\footnote{\url{https://github.com/vuejs}} and JavaScript, and the back end built with Python Flask\footnote{\url{https://github.com/pallets/flask}}. The system leverages the OpenAI \textit{gpt-4o} model for content generation.

\subsection{System Overview}
\par We meticulously designed \textit{CaseMaster}'s interface and interactions based on design goals identified through the formative study.
The probe guides users through two key stages: preparation and reflection. In the preparation stage, users start by viewing a patient list, organized from simple to complex cases (\textbf{DG1}). As they explore each patient's details, they interactively access resources and prepare their case using the SOAP format. \textit{CaseMaster} offers tailored materials and strategies for each step, with pre-set activities aligned with typical queries for the LLM (\textbf{DG4}). Users can regenerate content, edit drafts, and review their interaction history (\textbf{DG2}). After completing their preparation, users record and upload their solutions for assessment.
\par In the reflection stage, users transcribe their solutions, compare them with a reference answer, and receive a score based on a combination of an educator-provided score sheet and LLM feedback (\textbf{DG3}). \ouyangrevision{This stage is designed to support reflection and feedback integration, offering users structured contexts to review and refine their work (\textbf{DG5}). By integrating these stages with adaptable user controls, the system provides organized interaction with LLM-generated content (\textbf{DG4}) and is intended to guide users through training in a transparent and controllable manner (\textbf{DG6}).}

\subsection{Preparation Stage}
\begin{figure*}[h]
    \centering
    \includegraphics[width=\textwidth]{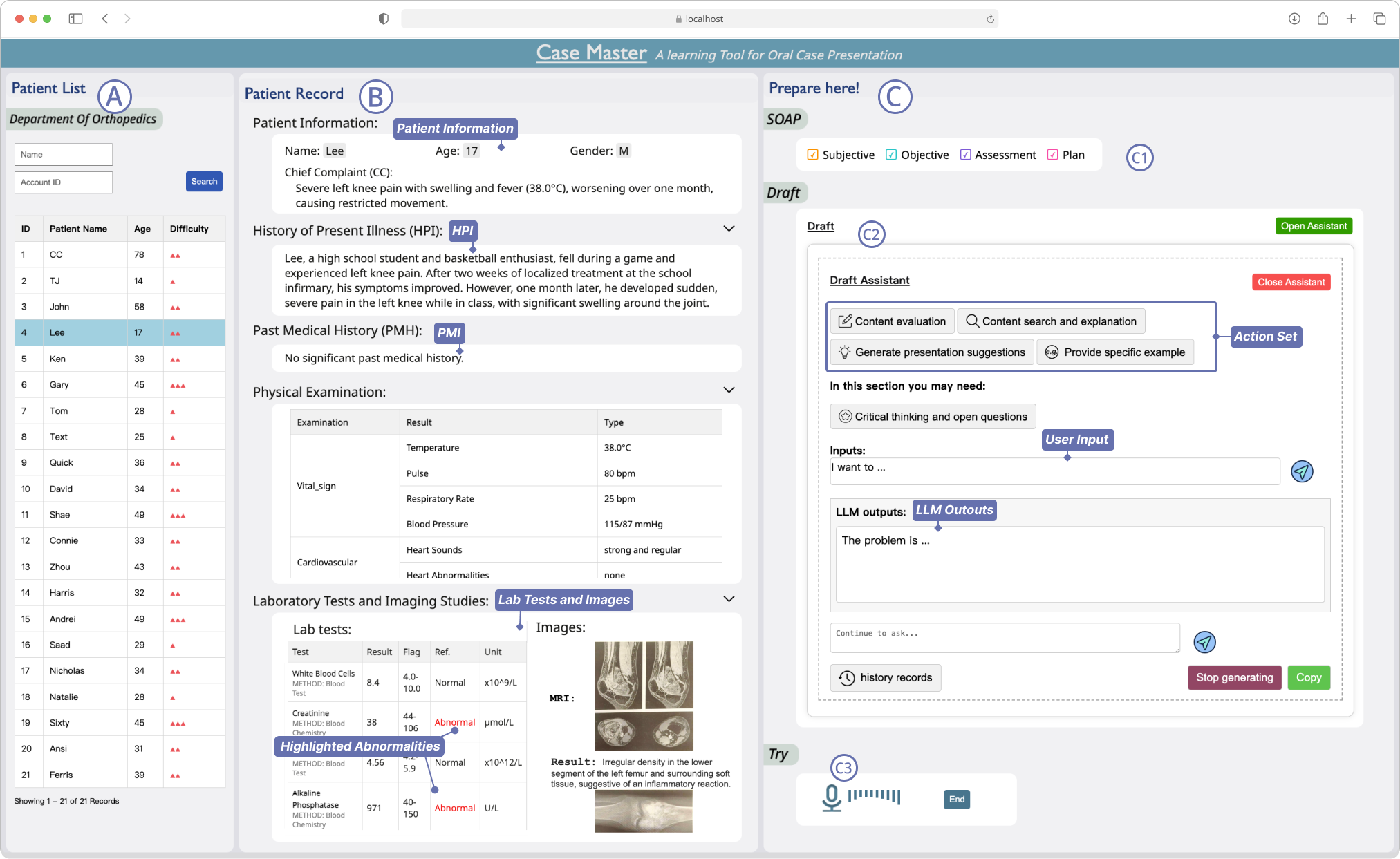}
    \caption{During the Preparation Stage, users can choose a patient from the (A) Patient List, review detailed information in the (B) Patient Record, and draft and refine their presentation within the (C) Preparation Panel. \ouyangre{The Preparation Panel includes LLM-powered assistance for content evaluation, suggestions, and generating specific examples to support users in crafting their presentation.}}
    \label{fig:prest}
\end{figure*}

\par In the Preparation Stage, \textit{CaseMaster} is organized into interactive components designed to streamline the creation and refinement of medical case presentations. The process begins with the \textit{Patient List} (\cref{fig:prest}-A), a structured overview where users can browse through patient records from the Department of Orthopedics. Each entry provides essential details, such as the patient's name, age, and a complexity indicator, serving as a quick reference for selecting a case. The table also features a search function, allowing users to quickly locate specific patients.

\par Upon selecting a patient, users are directed to the \textit{Patient Record} (\cref{fig:prest}-B). This section offers a comprehensive summary of the patient's medical history and current status, organized into distinct subsections. The \textit{Identifying Information} block includes demographic details, while the \textit{History of Present Illness (HPI)} provides a detailed narrative of the patient's symptoms, progression, and underlying conditions. The \textit{Laboratory Test} block lists diagnostic results, with abnormal values prominently highlighted for easy identification. Additionally, the \textit{Medication Use} block outlines the patient's current treatment regimen, including dosages, frequencies, and administration notes. This structured layout contributes to making all relevant information readily accessible for thorough case analysis.


\par The main focus of the case preparation process occurs in the \textit{Preparation Panel} (\cref{fig:prest}-C), where users draft their case presentations. This panel is structured around the SOAP framework, allowing users to customize their notes according to their specific needs. If a user deems a section, such as the \textit{Objective}, to be complete, they can mark it by clicking the corresponding checkbox (\cref{fig:prest}-C1). These checkboxes serve as visual cues, helping users structure their drafts effectively.

\par During each draft revision, users can click the ``Open the assistant'' button to activate the LLM assistant, which aids in refining their drafts (\cref{fig:prest}-C2). The assistant comprises the \textit{Action Set}, \textit{Input Area}, and \textit{Output Area}. The \textit{Action Set} includes a range of preset buttons that perform specific tasks, such as evaluating content, conducting content searches, and providing presentation advice. \ouyangrevision{\cref{sec:Pre_set} presents the activities implemented in \textit{CaseMaster}, which were informed by the design concepts developed during the co-design phase of the formative study. These activities provide structured guidance for completing tasks, helping users to focus on clinically relevant reasoning rather than on formatting or constructing queries for the LLMs. While these activities cover the areas identified through the co-design process, the system also allows for custom activities, supporting expansion to address additional tasks in future iterations.}

The \textit{Output Area} features a Markdown editor that displays the LLM's outputs after users interact with the \textit{Action Set} or make custom queries. Users can select content from the LLM's outputs and click the ``Copy'' button at the bottom of the assistant panel to copy it to the clipboard, allowing them to paste it anywhere within the current section's editor. The ``Stop Generating'' button halts the LLM's output, while the ``Close Assistant'' button hides the assistant. The ``History Record'' button provides access to previously generated content.

\par Users can draft key points for their final presentation and simulate the scenario by conducting an oral presentation by clicking the ``Voice'' button (\cref{fig:prest}-C3), much like TOEFL/IELTS practice~\cite{Zahedkazemi2015ConstructVO,Hadijah2018AnalisisKM}. 
\ouyangrevision{Although real-life simulations with actual partners provide a more immersive experience, this method offers a practical alternative. It facilitates users to practice on demand and focus on specific components of the presentation at their own pace.}



\subsection{Reflection Stage}
\label{sec:score}
\begin{figure*}[h]
    \centering
    \includegraphics[width=\textwidth]{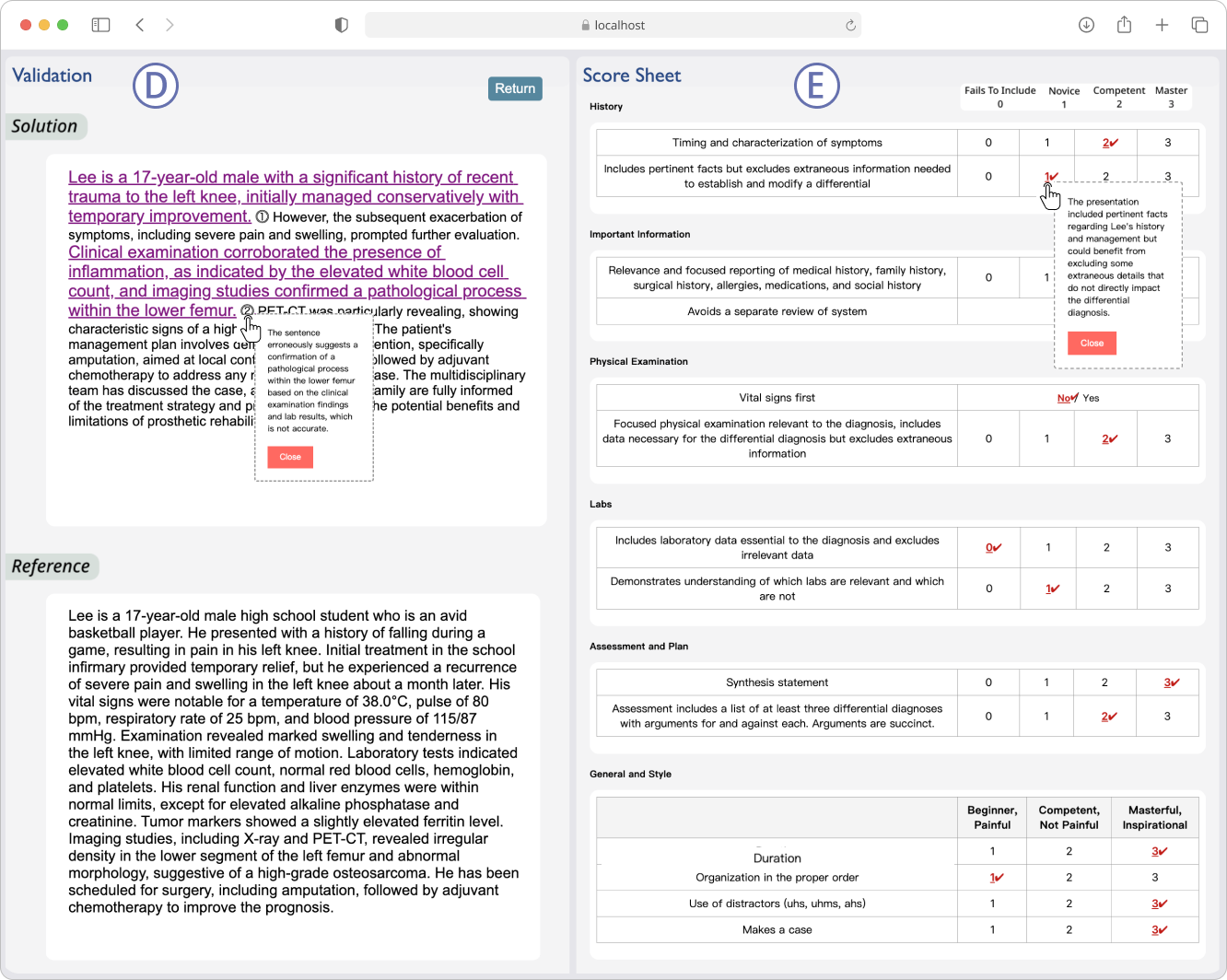}
    \caption{\ouyangre{During the Reflection Stage, users can compare their solutions in the (D) Validation panel, where highlighted discrepancies emphasize the differences between their solutions and the reference. The (E) Score Sheet displays LLM-generated scores across various evaluation criteria, with clickable red checkmarks offering detailed feedback.}}
    \label{fig:rest}
\end{figure*}

\par Upon completing the presentation, the probe transitions into the Reflection Stage, where \textit{CaseMaster} meticulously assesses the quality and thoroughness of the users' presentations. This stage consists of two main components: the \textit{Validation} (\cref{fig:rest}-D) and the \textit{Score Sheet} (\cref{fig:rest}-E). In the Validation section, users transcribe their responses and compare them to a reference answer. Discrepancies or missing elements are automatically highlighted, with distinct background colors indicating areas that need further review and understanding. Users can click on the highlighted sentence to view a detailed explanation of the identified mistakes.

\par In traditional educational settings, instructors often rely on detailed score sheets to evaluate students' video or audio submissions.  \ouyangrevision{The detailed score sheet is not a new framework separate from the SOAP structure (used in the Preparation stage), but rather functions as a granular rubric explicitly based on it. While SOAP provides the essential high-level organizational structure, the Reflection stage requires this finer-grained analysis to provide effective feedback.} To enhance efficiency and consistency, \textit{CaseMaster} leverages LLMs to automatically generate scores based on a comprehensive \textit{Score Sheet} and a reference answer. The LLM compares the student's work against the reference, identifying discrepancies and assessing the accuracy and completeness of each section.

\par In the \textit{History} section, the LLM assesses the timing and characterization of symptoms, as well as the inclusion of pertinent facts, using a $0$ to $3$ scale. The \textit{Important Information} section evaluates the relevance and focus of the medical history, ensuring unnecessary details are avoided, with scoring criteria that include both numerical scores and Yes/No evaluations. For the \textit{Physical Examination}, the LLM checks if vital signs are prioritized and whether the examination data is relevant to the diagnosis, again scoring from $0$ to $3$.
\par The \textit{Labs} section measures the inclusion and understanding of essential lab data, while \textit{Assessment and Plan} scores the clarity of the synthesis statement and the logic of differential diagnoses. The \textit{General and Style} section evaluates the overall organization, timing, and delivery of the presentation, scoring elements like duration, avoidance of filler words, and the strength of the argument on a scale from $1$ (Beginner, Painful) to $3$ (Masterful, Inspirational).


\par \ouyangrevision{The LLM compares student responses with the reference answer using predefined criteria, providing feedback to support thorough and objective evaluation.} On the \textit{Score Sheet}, scores in each category are highlighted in red with a checkmark-\checkmark, mimicking the real-world grading style of an educator. Users can click the checkmark to access detailed explanations for each score as provided by the LLM, including any detected mistakes.
\par Following the validation and assessment of the presentations, users can click the ``Return'' button to navigate back to the Preparation Stage, where they can either refine their existing presentations or select a new patient case to initiate a new round of presentations.


\subsection{Implementation Details}

\ouyangrevision{To support transparency and verifiability, we provide a detailed account of the technical implementation of CaseMaster’s LLM-assisted features, including data preparation, prompt design, model configuration, and validation of LLM-generated content.}

\subsubsection{Data Preparation and Anonymization}
\par Following methodologies outlined in several medical education studies~\cite{heiman2014improving,daniel2015teaching}, we collected data for \textit{CaseMaster} primarily from educational scenarios and select instructional videos. This approach, which includes recordings of simulated patient presentations and course examples, is grounded in proven educational practices. The scenarios are designed to closely mirror real-world clinical settings, providing students with realistic and challenging cases to enhance their diagnostic and communication skills. In total, we obtained $30$ oral case presentations, each accompanied by relevant patient medical records. These records were processed and organized into five key groups: \textit{Patient Information} \textit{(Name, Gender, Chief Complaint)}, 
\textit{History of Present Illness (HPI)}, \textit{Past Medical History (PMH)}, \textit{Physical Examination}, and \textit{Laboratory Tests and Imaging Studies}. Each case is also paired with a corresponding reference presentation answer.

\begin{table*}[h]
\centering
\caption{\ouyangrevision{Activity-Level evaluation of LLM-generated content in the preparation stage, rated by three senior medical educators.}}
\label{tab:activity_eval}
\resizebox{\linewidth}{!}{%
\begin{tabular}{>{\hspace{0pt}}m{0.206\linewidth}>{\hspace{0pt}}m{0.269\linewidth}>{\centering\hspace{0pt}}m{0.219\linewidth}>{\hspace{0pt}}m{0.344\linewidth}} 
\toprule
\textbf{Activity} & \textbf{Metric} & \textbf{Expert Average }\par{}\textbf{Rating} & \textbf{Notes} \\ 
\midrule
Search key knowledge points & Accuracy of key fact identification & 0.87 & High alignment; \par{}minor misses in subtle indicators. \\
Review medical literature & Correct literature reference \par{}and recommendation & 0.82 & LLM occasionally omits nuances in recent guidelines. \\
Check logic of content & Detection of contradictions & 0.85 & Reliable for basic inconsistencies; \par{}subtle clinical logic sometimes missed. \\
Assess reasonableness & Plausibility of recommendations & 0.80 & LLM may slightly overgeneralize for active patients. \\
Provide definitions & Correctness of medical term explanation & 0.95 & Very accurate; \par{}minor phrasing differences. \\
Provide specific example & Relevance and clarity of examples & 0.88 & High consistency; \par{}rare minor mismatches. \\
Explain examples in detail & Depth and accuracy of explanation & 0.83 & Occasionally lacks detail on rare complications. \\
Give presentation suggestions & Practicality and clarity of advice & 0.85 & Generally solid; \par{}small variance in emphasis of sequence. \\
\bottomrule
\end{tabular}
}
\end{table*}

\par The quality of the data has been validated by six medical educators. To facilitate student learning, each case's difficulty is categorized and scored into three levels: simple, intermediate, and advanced. This classification is primarily determined by case complexity, as well as factors like incidence rate and required reasoning skills, offering a structured path for students to progressively enhance their diagnostic abilities. We obtained IRB approval from the Research Ethics Committee. \ouyangrevision{All data were anonymized through a rule-based de-identification pipeline, in which patient names, dates of birth, and other direct identifiers were replaced with randomly generated codes, and any contextual details that could reveal identity were carefully removed or generalized.} As a probe, we initially selected patients from the same department to minimize variations in focus across specialties and reduce differences in emphasis during oral case presentations. We may expand the dataset to include patients from other departments in the future.

\subsubsection{Prompt Structure}
\par \ouyangrevision{In the Preparation Stage, the LLM functions are implemented through a set of modular prompt templates aligned with the activity types listed in \cref{sec:Pre_set}. Each template explicitly defines the activity's objective, outlines the expected reasoning steps, and specifies the input and output formats, while including concise input–output exemplars that illustrate the intended transformation and level of detail. These exemplars are appended to the end of each prompt to stabilise model behaviour and constrain verbosity. For the Reflection Stage, scoring prompts direct the LLM to assume the role of an expert educational evaluator, compare the student responses with the reference answer, and return a structured JSON object in accordance with the detailed rubric. The scoring exemplar illustrates a correct mapping between student responses and rubric items, helping to minimize mismatches and prevent the model from generating unsupported deductions. Two detailed example prompts are provided in \cref{sec:pro_example}.}

\subsubsection{Model Configuration and Temperature Tuning}
\ouyangrevision{\textit{CaseMaster} uses the OpenAI \textit{gpt-4o} model via a Flask backend. We empirically determined temperature settings for different tasks via pilot testing on three representative cases. For evaluative tasks (Reflection Stage scoring), temperatures from 0.1 to 0.5 were tested, and \textbf{0.2} was selected to maximize alignment with expert judgments while minimizing output variability. For generative tasks in the Preparation Stage (e.g., \textit{Provide specific example}), temperatures from 0.3 to 0.6 were evaluated, and \textbf{0.5} was chosen to balance creativity and reliability. Output length is constrained to maintain concise responses aligned with OCP training goals. Each request follows a structured system--user--assistant format, with patient data and task instructions explicitly injected to minimize context drift. These temperature settings were chosen to balance stability for assessment tasks and expressiveness for content generation, aiming to support the instructional goals of OCP training.}

\subsubsection{Validation of LLM-Generated Content}
\ouyangrevision{To validate the LLM-generated content, we conducted two evaluations corresponding to the system's primary stages: one for the LLM-assisted activities (Preparation Stage) and one for the automated scoring (Reflection Stage).}

\par \textit{Activity-Level Evaluation for Preparation Stage.} 
\ouyangrevision{We evaluated LLM-generated content using four randomly selected patient cases, focusing on the set of core pedagogical activities detailed in \cref{sec:Pre_set} (e.g., identifying key knowledge points, reviewing relevant literature, and checking content logic). For each activity, three experienced medical educators collaboratively pre-designed the corresponding LLM query and evaluation metrics to align with core pedagogical goals and expected reasoning steps. The educators then rated each activity's output on a 0–1 scale, recording scores to two decimal places, with higher values indicating better performance. The final scores were averaged across the three raters (\cref{tab:activity_eval}). Overall, the model demonstrated relatively reliable performance across all metrics. However, minor gaps were observed, such as subtle omissions in key facts, an occasional lack of nuance in literature references, or limited depth in complex explanations. This suggests that students should be guided to review the content critically rather than accept it unconditionally.}

\begin{table*}[h!]
\centering
\caption{\ouyangre{The participant demographics and characteristics in the experiment include gender, age, academic level, specialization, and experience with OCP. Experience levels are categorized from lowest to highest as follows: Limited OCP Experience, OCP Training \& Peer Mentor, and OCP Training Assistant. Additionally, the data highlights each participant's frequency of LLM usage.}}

\resizebox{\linewidth}{!}{
\begin{tabular}{ccccccc}
\hline
\textbf{ID} & \textbf{Gender} & \textbf{Age} & \textbf{Year} & \textbf{Specialization} & \textbf{OCP Experience} & \textbf{Frequency of LLM Usage} \\ \hline
P1  & M & 23 & Graduate      & Orthopedics      & OCP Training \& Peer Mentor   & Daily        \\
P2  & M & 24 & Graduate      & Orthopedics      & OCP Training \& Peer Mentor   & Weekly       \\ 
P3  & M & 25 & Graduate      & Orthopedics      & Limited OCP Experience        & Occasionally \\ 
P4  & M & 23 & Graduate      & Rehabilitation   & Limited OCP Experience        & Weekly       \\
P5  & M & 24 & Graduate      & Orthopedics      & OCP Training \& Peer Mentor   & Weekly       \\ 
P6  & M & 24 & Graduate      & Rehabilitation   & OCP Training Assistant        & Occasionally \\
P7  & M & 22 & Undergraduate & Orthopedics      & Limited OCP Experience        & Weekly       \\
P8  & M & 25 & Graduate      & Orthopedics      & Limited OCP Experience        & Daily        \\
P9  & F & 29 & Graduate      & Rehabilitation   & OCP Training Assistant        & Occasionally \\
P10 & F & 23 & Undergraduate & Rehabilitation   & OCP Training \& Peer Mentor        & Weekly       \\ 
P11 & F & 22 & Undergraduate & Orthopedics      & Limited OCP Experience        & Daily        \\ 
P12 & M & 23 & Graduate      & Orthopedics      & Limited OCP Experience      & Daily        \\ \hline
\end{tabular}
}
\label{tab:participants}
\end{table*}

\par \textit{Scoring Evaluation for Reflection Stage.}
\ouyangrevision{To validate the LLM-based scoring, we used the same four patient cases along with eight corresponding student-generated presentations. These presentations were independently scored by three medical educators as well as by \textit{CaseMaster}. Inter-rater reliability between the LLM-generated scores and the average expert scores was assessed using the Intraclass Correlation Coefficient (ICC = 0.88), indicating high agreement. We also manually reviewed all 112 LLM-generated scoring items for qualitative errors, identifying three minor issues (e.g., incorrectly marking a correctly reported lab result as missing). Taken together, these results indicate that the scoring is generally reliable, while also highlighting the importance of having students review the reference answers themselves during the training process.}




\section{EVALUATION OVERVIEW}

\par \ouyangrevision{To investigate how \textit{CaseMaster} supports medical students in OCP training, we conducted two complementary user studies between March and May 2025, gathering feedback from both students and educators.} The first study, a experimental user study, involved university medical students who used \textit{CaseMaster} alongside a baseline tool to complete an oral case presentation. We analyzed their perceptions of the usability of \textit{CaseMaster} and their comparison of it with the baseline tool. The second study engaged five experienced medical educators who evaluated \textit{CaseMaster}, providing insights into how they interact with and perceive the tool. Through qualitative interviews, these educators offered constructive feedback, contributing valuable perspectives on \textit{CaseMaster}'s potential to enhance medical education. \ouyangrevision{Together, these studies provide complementary evidence, integrating both student and educator perspectives to inform the design and use of \textit{CaseMaster}.}
These two studies aim to answer the following research questions (RQs):

\begin{itemize}
\item \textbf{RQ1:} How would \textit{CaseMaster} affect the outcome of oral case presentations?
\item \textbf{RQ2:} How would \textit{CaseMaster} affect the process of delivering oral case presentations?
\item \textbf{RQ3:} How would users perceive \textit{CaseMaster} as a tool for training in oral case presentations?
\end{itemize}
We also provide a System Walkthrough in \cref{sec:walkthrough}.

\section{Evaluation Study 1: Experiment}
To evaluate the user experience provided by \textit{CaseMaster}, we conducted a experimental user study with 12 university medical students who used the tool to practice delivering an oral case presentation. This study focuses on students' perspectives, aiming to evaluate if they find \textit{CaseMaster} both intuitive and effective for OCP training.
\subsection{Method}
\subsubsection{Participants}
\par We recruited $12$ medical students (P1-P12, three females, nine males; age: $\text{Mean} = 23.92$, $\text{SD} = 1.80$) through word-of-mouth at a local hospital, facilitated by our collaborating team (\cref{tab:participants}). The group consisted of three $5^{th}$-year undergraduates and nine graduate students, all majoring in medicine with prior training in OCP as part of their curriculum. At the time of the study, all participants were undergoing internships in rehabilitation or orthopedic departments at the hospital.
All participants expressed a strong interest in exploring digital tools to enhance their OCP skills ($\text{Mean} = 4.08$, $\text{SD} = 0.64$; 1 = no interest at all, 5 = very interested). \ouyangre{Additionally, four participants reported daily use of LLMs, such as ChatGPT and MedicalGPT, and felt quite familiar with their features, finding them helpful for work and study. Similarly, five participants who used them weekly also expressed comfort in using LLMs to assist with their tasks.} IRB approval was obtained prior to the start of the study.

\subsubsection{Experiment Design}
\label{sec:ex_design}
\par The goal of this study is to quantitatively assess the effectiveness and user experience of \textit{CaseMaster} compared to a baseline system. It should be noted that this comparison is not to demonstrate one system's superiority but to gain insights into how each influences user behaviors and mindsets.
\begin{figure*}[h]
    \centering
    \includegraphics[width=\textwidth]{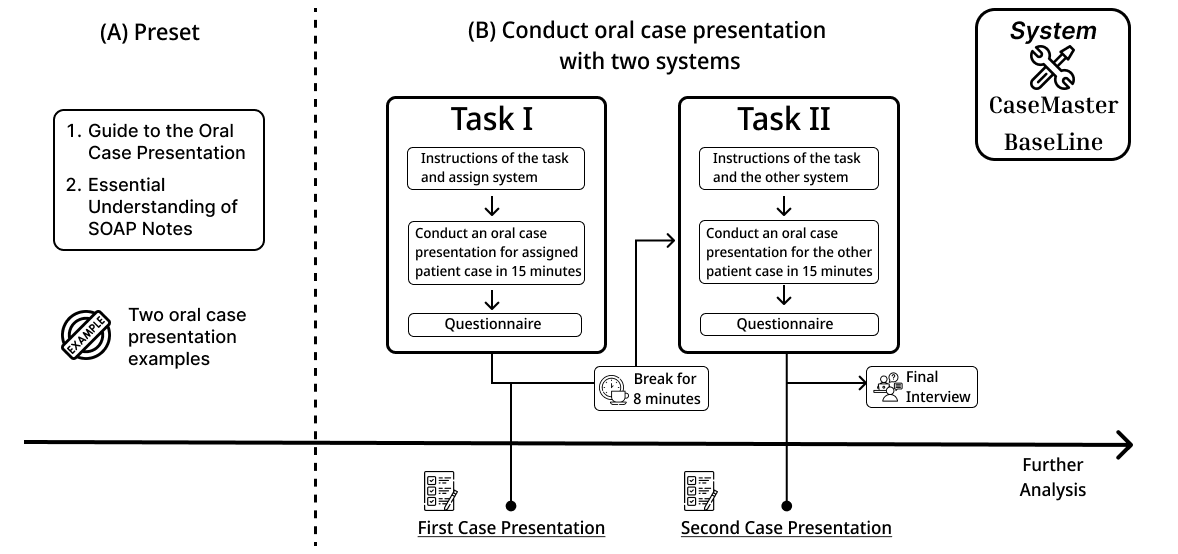}
    \caption{\ouyangre{The procedure for the user study comparing \textit{CaseMaster} with a baseline system was conducted remotely via Zoom. Participants completed two OCP tasks using pre-study materials, followed by independent preparation, completion of feedback questionnaires, and a final interview discussion.}}
    \label{fig:userstudy}
\end{figure*}

\par \textbf{Baseline Condition.} The baseline condition replicates the traditional, self-directed training workflow typical for medical students. Participants were given standard educational tools to support guided practice and independent exploration: the Google search engine for information gathering, a markdown editor for preparation, and optional access to ChatGPT for assistance. Crucially, to prevent the introduction of structured support comparable to \textit{CaseMaster}, no prompts, templates, or task decomposition were provided for ChatGPT. In practice, participants used ChatGPT intermittently during case preparation, most frequently for clarifying information or verifying specific reasoning steps. Usage varied, with most consulting it more than three times per case and a minority using it minimally. This condition imposed no constraints on external resources, establishing an open reference point for comparison with \textit{CaseMaster}.


\par \textbf{CaseMaster Condition.} In the \textit{CaseMaster} condition, participants are allowed to use the Google search engine to gather information and resources relevant to their OCPs. However, access to the ChatGPT web app or MedicalGPT is restricted. This limitation is intended to guarantee that the study focuses solely on the capabilities of \textit{CaseMaster}, without external influences from other AI tools.

\par \textbf{Task-System Assignment.} During the recruitment survey, participants are asked to select the patient cases they are most interested in exploring. Based on their responses, we identify two cases that receive unanimous interest. To ensure balance, we use a Latin Square design to counterbalance the tasks, assigning three participants to each of the four potential sequences. The sequences are as follows: three participants follow the sequence ``\textit{CaseMaster} - Case I -> Baseline - Case II'', three follow ``Baseline - Case II -> \textit{CaseMaster} - Case I'',  three follow ``\textit{CaseMaster} - Case II -> Baseline - Case I'', and the remaining three follow ``Baseline - Case I ->  \textit{CaseMaster} - Case II''.

\subsubsection{Tasks and Procedure}
\par \cref{fig:userstudy} outlines the procedure and tasks of our user study, which was conducted remotely via Zoom. Each participant was assigned two tasks. Prior to the study, participants were provided with instructional materials, including a guide to OCP, essential knowledge on SOAP format notes, and two example videos of OCPs. They were instructed to spend 10-15 minutes reviewing these materials before completing the assigned tasks.

\par On the day of the oral case presentation tasks, participants were grouped in sets of three according to their task-system assignment. We provided a brief introduction to relevant examples and highlighted the essential elements of an effective OCP, based on the materials distributed the previous day.

\par For each task, we begin by introducing the task and demonstrating the assigned probe to the participants. We assist them in setting up their operating environments, ensuring that the necessary webpages-whether for the baseline system or \textit{CaseMaster}, along with Google and instructional materials-are open and ready. Participants then work independently on the OCP preparation, with $15$ minutes allocated per task. They are allowed to finish early or take a few extra minutes if needed. After completing each task, participants fill out a questionnaire to share their experiences and impressions of the system used, followed by a brief 8-minute interview. An 8-minute break is provided between tasks for participants to rest. Once both tasks are completed, we conduct a final semi-structured interview, where participants compare the systems, reflect on the content produced, discuss \textit{CaseMaster}'s features, and offer suggestions for improvement. Each participant spends approximately $90$ minutes in the experiment and receives a \$20 Amazon gift card as compensation.

\subsubsection{Measurements}
\par We employ a standard 7-point Likert scale (1 - strongly disagree, 7 - strongly agree) to measure case presentation quality, the experience during the presentation process, and participants' perceptions of the probe.

\par \textbf{RQ1: Case Presentation Outcome.} 
\ouyangrevision{To support a rigorous and reliable evaluation of presentation quality, two educators with three years of experience in OCP training were engaged as raters.} They independently reviewed the presentations in a randomized order, assessing only the content extracted from video recordings of the case presentations. The evaluation criteria were adapted from the aforementioned score sheet (\cref{sec:score}), and included the following six aspects: \textit{characterization of symptoms}, \textit{focus of medical history}, \textit{physical examination}, \textit{lab relevance}, \textit{differential diagnosis}, and \textit{presentation organization and style}. By using a 7-point scale, we aimed to avoid oversimplification and better reflect even subtle differences in presentation quality in the ratings.

\par \ouyangre{To maintain consistency and support a shared understanding of the evaluation criteria, both raters participated in a training session before the independent assessment phase. In this session, they reviewed five sample presentations together, discussing how to apply the evaluation criteria and resolving any ambiguities. This collaborative training \ouyangrevision{facilitated} alignment between the raters in their approach and interpretation of the criteria, further enhancing the reliability of the evaluation. Finally, after completing their independent assessments, inter-rater reliability, measured using Cohen's Kappa metric \cite{mchugh2012interrater}, demonstrated strong agreement between the two raters ($\kappa = 0.86$), confirming the consistency and reliability of the evaluation process. The raters then met to discuss their assessments, reach a consensus, and address any remaining discrepancies, ensuring the final ratings accurately reflected the presentation quality.} 

\par \textbf{RQ2: Case Presentation Process.} Based on the NASATLX survey~\cite{hart1988development}, we asked six questions to assess the workload during the case presentation process, covering \textit{Mental Demand}, \textit{Physical Demand}, \textit{Temporal Demand}, \textit{Performance}, \textit{Effort}, and \textit{Frustration}. \ouyangre{Participants were asked: ``\textit{How mentally demanding did you find the entire task using the system?}'' ``\textit{How physically demanding was the entire task with the system?}'' ``\textit{Did you feel pressured by time constraints during the process with the system?}'' ``\textit{How successful do you think the task was using the system?}'' ``\textit{How much overall effort did you invest in the task with the system?}'' and ``\textit{How frustrated did you feel while completing the task with the system?}'' These questions enabled a direct comparison of the perceived workload between both systems.}
\ouyangre{In addition to the quantitative ratings, follow-up interviews were conducted to gather qualitative data. Participants were encouraged to discuss how the probe impacted their workload in each of these areas. They were asked to reflect on any increase or decrease in their mental, physical, or emotional demands, providing us with a deeper understanding of their experiences. For instance, participants who reported high frustration levels were prompted to elaborate on the reasons for their feelings—whether they stemmed from difficulty using the presentation tool, anxiety about delivering the presentation, or external factors like time pressure.}

\par \textbf{RQ3. Perceptions towards \textit{CaseMaster}.} Participants completed a post-study questionnaire to evaluate their experiences, which included a System Usability Scale (SUS)~\cite{brooke2013sus} using Likert scales. The questionnaire aimed to assess the system's usefulness, ease of use, satisfaction and Likelihood of future use. \ouyangre{\textit{Usefulness} was measured by the items:``\textit{I think that I would like to use this system frequently}'' and ``\textit{I would imagine that most people would learn to use this system very quickly}''; \textit{Ease of Use} was assessed by the items: ``\textit{I thought the system was easy to use,}'' ``\textit{I found the various functions in this system were well integrated,}'' ``\textit{I felt very confident using the system,}'' and ``\textit{I needed to learn a lot of things before I could get going with this system}''; \textit{Satisfaction} was measured by the items: ``\textit{I think that I would need the support of a technical person to be able to use this system,}'' ``\textit{I found the system very awkward to use,}'' and ``\textit{I thought there was too much inconsistency in this system}'' and \textit{Likelihood of Future Use} was measured by ``\textit{I think that I would like to use this system frequently.}''} We averaged the ratings of multiple questions as the final score for each aspect. In addition, we adapted five questions from Jian's Trust Scale~\cite{jian2000foundations} to examine participants' trust in the system, focusing on aspects such as system vigilance, potential negative effects, ethical standards, user commitment, and output reliability, as shown in \cref{sec:system_trust}.
During the interviews, participants were encouraged to discuss their experiences and preferences for each tool (i.e., search engines, the ChatGPT web app, and \textit{CaseMaster}), as well as various components of \textit{CaseMaster} (e.g., different pre-set activities and features). We also explored their reasons for trusting or distrusting the system in more detail.

\subsubsection{Data Analysis}
\label{sec:dataanalysis}
All study sessions were recorded and transcribed. Data collection included server-side logs, screen recordings, and interviews. For the quantitative analysis, we employed a 7-point Likert scale to assess outcomes related to OCP performance (RQ1), the OCP process (RQ2), and user perceptions of \textit{CaseMaster} (RQ3). We opted for non-parametric statistical methods given the ordinal nature of Likert-scale responses and the small sample size. Specifically, to compare \textit{CaseMaster} with the baseline method, we conducted a Wilcoxon signed-rank test~\cite{woolson2005wilcoxon} as part of our experiment. \ouyangrevision{The results are reported with $W$ denoting the Wilcoxon test statistic, representing the sum of signed ranks of differences between paired observations, and $p$ indicating the corresponding significance level. Medians ($M$) and interquartile ranges (IQR) are provided to summarize the central tendency and variability for each condition. Given that multiple pairwise comparisons were conducted within each research question, we applied a Bonferroni correction to control for Type I error. Specifically, for $n$ comparisons, raw p-values were multiplied by $n$. These adjusted p-values (reported) were considered significant if they were less than the conventional 0.05 threshold.} 
\par \ouyangrevision{For these tests, we formulated an explicit null hypothesis for each research question: that there is no difference between \textit{CaseMaster} and the baseline method ($H_0$), against the alternative hypothesis that there is a difference ($H_1$).}

\par \ouyangre{For the qualitative data collected, two authors independently reviewed the transcript for each participant, distilling key insights. The union of these emergent insights was then used to create affinity diagrams, which helped synthesize and organize observations across the interviews. These diagrams were inspected, labeled, and discussed with all authors, revealing key themes and patterns. The content was further analyzed to categorize and prioritize themes, merging or removing overlapping clusters as needed. After finalizing the diagrams, two authors independently verified all findings against the original transcripts, ensuring consistency and identifying no discrepancies. We chose affinity diagrams over grounded theory for several reasons. This method is commonly used in HCI and interaction design practice~\cite{harboe2015real,lucero2015using}. Moreover, our focus was not on developing a theoretical framework, but on evaluating the design and application of new tools for OCP training in the context of a user study. By directly engaging medical students with \textit{CaseMaster}, we aimed to assess their experience using the tool to practice oral case presentations. 
}

\subsection{Results}
\subsubsection{Case Presentation Outcome (RQ1)}

\par
\begin{figure}[h]
    \centering
    \includegraphics[width=\linewidth]{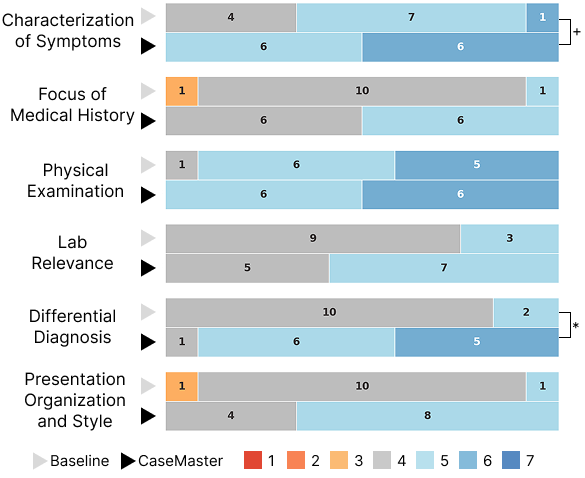}
    \caption{The results for RQ1, regarding the Case Presentation Outcome, were evaluated by experienced teachers across six aspects, comparing the baseline and \textit{CaseMaster}. Significance levels are indicated as follows: ***: $p < 0.001$, **: $p < 0.01$, *: $p < 0.05$, +: $p < 0.1$, with all $p$-values adjusted using the Bonferroni correction. Numbers in each cell denote the number of participants assigning the corresponding rating.}
    \label{fig:result_1}
\end{figure}

 \cref{fig:result_1} presents the results of the oral case presentation outcomes. \ouyangrevision{After applying a Bonferroni correction for six comparisons, the only subcomponent that remained statistically significant was the clarity of differential diagnoses (\textit{CaseMaster}: $M = 5.00$, $IQR = 1.00$; baseline: $M = 4.00$, $IQR = 0.00$; $W = 3.50$, $p_{\text{adj}} = 0.042$), providing evidence to reject the null hypothesis and supporting $H_1$ that \textit{CaseMaster} enhances the quality of differential diagnoses for \textbf{RQ1}. In contrast, other areas did not reach statistical significance after the correction, though they still showed favorable trends for \textit{CaseMaster}. For instance, the alignment of physical examination content showed no significant difference between \textit{CaseMaster} ($M = 5.50$, $IQR = 1.00$) and the baseline system ($M = 5.00$, $IQR = 1.00$; $W = 10.50$, $p_{\text{adj}} = 1.0$). Presentations also showed trends toward better focus on critical medical history points (\textit{CaseMaster}: $M = 4.50$, $IQR = 1.00$; baseline: $M = 4.00$, $IQR = 0.00$; $W = 3.50$, $p_{\text{adj}} = 0.348$) and in incorporating relevant laboratory data (\textit{CaseMaster}: $M = 5.00$, $IQR = 1.00$; baseline: $M = 4.00$, $IQR = 0.25$; $W = 3.50$, $p_{\text{adj}} = 0.612$). Moreover, \textit{CaseMaster} presentations tended to exhibit better organizational structure and logical sequencing (\textit{CaseMaster}: $M = 5.00$, $IQR = 1.00$; baseline: $M = 4.00$, $IQR = 0.00$; $W = 4.50$, $p_{\text{adj}} = 0.126$) and more effective symptom characterization (\textit{CaseMaster}: $M = 5.50$, $IQR = 1.00$; baseline: $M = 5.00$, $IQR = 1.00$; $W = 5.00$, $p_{\text{adj}} = 0.078$), though these differences did not meet the Bonferroni-adjusted significance threshold.}
Participants then highlighted how \textit{CaseMaster} positively impacted the quality of their case presentations, as detailed below.

\textbf{\ouyangrevision{Effects of LLM Assistance on Case Presentation Quality.}}  
The majority of participants (9 out of 12) reported that the preparation stage in \textit{CaseMaster} noticeably improved the quality of their drafts. \ouyangrevision{They favored \textit{CaseMaster}'s preset activities, which provided clear and structured guidance. While overall LLM usage was higher with \textit{CaseMaster} than with the baseline tool (ChatGPT), individual patterns still reflected personal tendencies, with some participants consulting the LLM more frequently than others.}


\par \textit{\ouyangrevision{Support for novices.}}  
Participants with limited OCP experience (e.g., P3, P4, P7, P11) frequently reported that \textit{CaseMaster} clarified their thinking and expanded their diagnostic reasoning. For example, P3 noted: \textit{"Features like giving and explaining examples really help me understand the case better and dive deeper into it."} Similarly, P7 highlighted the tool’s broader perspective: \textit{"Instead of focusing on just one symptom, it takes a broader view and supports it with evidence-based examples."} Moreover, two novices (P4, P7) described the LLM’s explanations as "both engaging and helpful in identifying their mistakes," emphasizing the value of interactive, generative guidance.

\par  \textit{\ouyangrevision{Workflow preferences of experienced physicians.}} 
\ouyangrevision{In contrast, some experienced participants found the structured approach less compatible with their established workflows. P9, an OCP Training Assistant, explained: \textit{"I prefer structuring my thoughts visually on slides, as it helps me organize content more clearly and see the overall flow of the presentation."} This feedback suggests opportunities for personalization, like integrating LLM support into slide-based preparation.}

\par  \textit{\ouyangrevision{Widely Valued Features.}} 
Despite differences in workflow preferences, the reflection stage was widely appreciated by both novice and experienced users (P2, P4, P5, P7, P10). While novices benefit more from generative guidance (i.e., help creating the draft), all participants valued the LLM’s rapid evaluative feedback on their solutions. An experienced participant (P6, OCP Training Assistant) recommended a scaffolded approach to feedback: \textit{"The probe doesn't need to give the answers right away. It could start by pointing out what's wrong, then later show the full solution, kind of like using a step-by-step approach."} This indicates a desire for more incremental guidance that supports deeper understanding.

\begin{figure}
    \centering
    \includegraphics[width=\linewidth]{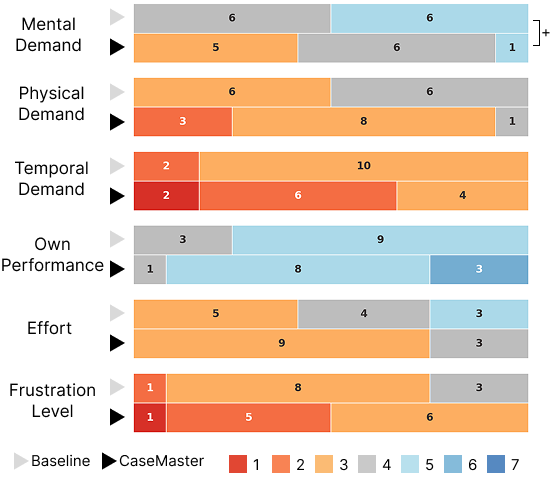}
    \caption{\ouyangrevision{The results for RQ2, which assess task load, are based on pairwise comparisons between the baseline and \textit{CaseMaster}. Significance levels are indicated as follows: ***: $p < 0.001$, **: $p < 0.01$, *: $p < 0.05$, +: $p < 0.1$, with all $p$-values adjusted using the Bonferroni correction.} Numbers in each cell denote the number of participants assigning the corresponding rating.}
    \label{fig:result_2}
\end{figure}

\subsubsection{Case Presentation Process (RQ2)}
\par \cref{fig:result_2} presents participants’ perceived task workload during case presentations with \textit{CaseMaster} versus the baseline system. \ouyangrevision{After applying a Bonferroni correction for six comparisons, none of the workload subcomponents reached the adjusted significance threshold. Specifically, participants reported lower mental demand (\textit{CaseMaster}: $M = 4.00$, $IQR = 1.00$; baseline: $M = 4.50$, $IQR = 1.00$; $W = 4.50$, $p_{\text{adj}} = 0.078$), physical demand (\textit{CaseMaster}: $M = 3.00$, $IQR = 0.25$; baseline: $M = 3.50$, $IQR = 1.00$; $W = 3.50$, $p_{\text{adj}} = 0.198$), temporal demand (\textit{CaseMaster}: $M = 2.00$, $IQR = 1.00$; baseline: $M = 3.00$, $IQR = 0.00$; $W = 4.50$, $p_{\text{adj}} = 0.126$), frustration (\textit{CaseMaster}: $M = 2.50$, $IQR = 1.00$; baseline: $M = 3.00$, $IQR = 0.25$; $W = 2.50$, $p_{\text{adj}} = 0.282$), effort (\textit{CaseMaster}: $M = 3.00$, $IQR = 0.25$; baseline: $M = 4.00$, $IQR = 1.25$; $W = 8.00$, $p_{\text{adj}} = 0.420$), or perceived task performance (\textit{CaseMaster}: $M = 5.00$, $IQR = 0.25$; baseline: $M = 5.00$, $IQR = 0.25$; $W = 2.00$, $p_{\text{adj}} = 0.774$). Accordingly, we cannot reject the null hypothesis for \textit{RQ2} (no difference in perceived task workload) in any of these dimensions, though the raw trends suggest participants perceived somewhat lower workload when using \textit{CaseMaster}.
} Below, we provide further insights into the task workload from the participants.
\par \textbf{\ouyangrevision{Perspectives on the Workload of the Case Presentation Process.}}
Our controlled study demonstrates that \textit{CaseMaster} effectively reduces task workload compared to the baseline interface. Participants noted that the structured control over certain LLM outputs allowed them to consult the ``assistant'' for clarification when uncertain, avoiding confusion. As P11 remarked, ``\textit{..the probe gives me different content to choose [from], so I can pick what fits my needs best. Sure, it takes a little [more] effort, but before, I only had one basic option to work with}''. On the other hand, P9 shared their experience with the baseline interface, saying, ``\textit{When I saw that the results from the generative model weren't quite what I needed, I kept tweaking [it] and refining the outputs. Of course, that ended up taking more time}''.



\subsubsection{Perceptions towards CaseMaster (RQ3)}

\begin{figure}
    \centering
    \includegraphics[width=\linewidth]{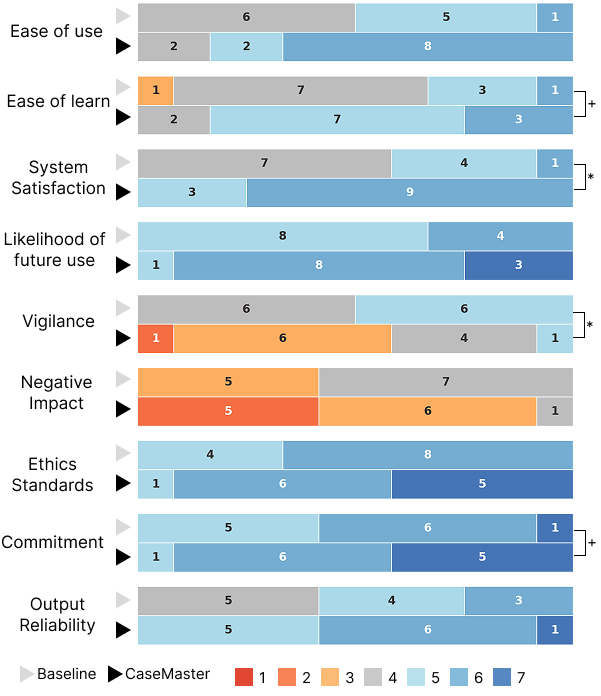}
    \caption{\ouyangrevision{The results for RQ3, which evaluate system usability and participants' trust, are based on pairwise comparisons between the baseline and \textit{CaseMaster}. Significance levels are indicated as follows: ***: $p < 0.001$, **: $p < 0.01$, *: $p < 0.05$, +: $p < 0.1$, with all $p$-values adjusted using the Bonferroni correction.} Numbers in each cell denote the number of participants assigning the corresponding rating.}
    \label{fig:result_3}
\end{figure}
\par \cref{fig:result_3} shows participants' usability and trust ratings for \textit{CaseMaster} compared to the baseline system. \ouyangrevision{After Bonferroni correction for nine comparisons, participants rated \textit{CaseMaster} higher across multiple aspects, showing positive trends. On usability, \textbf{System Satisfaction} ($M = 6.00$, $IQR = 0.25$; baseline: $M = 4.00$, $IQR = 1.00$; $W = 4$, $p_{\text{adj}} = 0.027$) approached significance, supporting $H_1$ for this dimension. Other usability dimensions, including \textbf{Ease of Use} ($M = 6.00$, $IQR = 1.00$; baseline: $M = 4.50$, $IQR = 1.00$; $p_{\text{adj}} = 0.468$), \textbf{Ease of Learning} ($M = 5.00$, $IQR = 0.25$; baseline: $M = 4.00$, $IQR = 1.00$; $p_{\text{adj}} = 0.063$), and \textbf{Likelihood of Future Use} ($M = 6.00$, $IQR = 0.25$; baseline: $M = 5.00$, $IQR = 1.00$; $p_{\text{adj}} = 0.180$), also trended in favor of \textit{CaseMaster} but did not reach the adjusted significance threshold, thus retaining $H_0$. Regarding trust, \textbf{Vigilance} ($M = 3.00$, $IQR = 1.00$; baseline: $M = 4.50$, $IQR = 1.00$; $W = 5$, $p_{\text{adj}} = 0.045$) showed a notable reduction relative to baseline, indicating a directional trend but still above the corrected threshold. Other trust dimensions, including \textbf{Negative Impact} ($M = 3.00$, $IQR = 1.00$; baseline: $M = 4.00$, $IQR = 1.00$; $W = 2$, $p_{\text{adj}} = 0.189$), \textbf{Ethical Standards} ($M = 6.00$, $IQR = 1.00$; baseline: $M = 6.00$, $IQR = 1.00$; $W = 3.50$, $p_{\text{adj}} = 0.297$), \textbf{Commitment to User} ($M = 6.00$, $IQR = 1.00$; baseline: $M = 6.00$, $IQR = 1.00$; $W = 6.50$, $p_{\text{adj}} = 0.081$), and \textbf{Reliability of Outputs} ($M = 6.00$, $IQR = 1.00$; baseline: $M = 5.00$, $IQR = 1.25$; $W = 3.50$, $p_{\text{adj}} = 0.180$), also favored \textit{CaseMaster} but did not reach the Bonferroni-adjusted significance threshold.} The interviews revealed that participants actively shared their insights on \textit{CaseMaster}'s usability and voiced their trust or concerns about the generated content. This feedback is summarized below.

\par \textbf{\ouyangrevision{Perceived benefits of \textit{CaseMaster} for OCPs.}}
In the experiment, participants appreciated that \textit{CaseMaster} provides inspiring content for case presentations. Specifically, all participants found the ``Provide an Example in Detail'' feature to be well-designed and highly useful for a more comprehensive exploration of the case. \ouyangrevision{Notably, this feature was valued not only by novices but also by experienced participants.} For instance, P2 mentioned, ``\textit{...Sometimes when I'm working on a patient case, I come up with ideas but I'm not always sure if they're right. The system provides detailed examples that match my thoughts, which really helps me clear up my thinking and dive deeper into the case.}''


\par \ouyangrevision{\textbf{Perceptions of \textit{CaseMaster}'s reliability.}}
In the within-subject study, eight participants reported higher confidence in the content generated by \textit{CaseMaster} compared to the baseline system. For example, P4 (a student with `Limited OCP Experience') and P10 (an `OCP Training \& Peer Mentor') both felt that the preset activities and structured format contributed to \textit{CaseMaster} appearing more professional and trustworthy than the open-ended baseline system.
\ouyangrevision{However, trust was not universal and appeared linked to participants' LLM familiarity and personal experience:}
\begin{itemize}

    \item \ouyangrevision{\textit{Impact of familiarity:} Participants who interacted with LLMs more regularly (either `Daily' or `Weekly') generally had more trust in the system. This highlights a challenge in demonstrating system logic to less-frequent LLM users.}
    
    \item \ouyangrevision{\textit{Emerging concerns:} P9 (an `OCP Training Assistant' with `Occasional' LLM use) expressed a typical expert concern about ``black box'' systems, requiring transparency to build trust: ``\textit{I may have some understanding of how prompts work, but I would only trust the output result if I can see all the details of the prompts.}''}

\end{itemize}


\par \ouyangrevision{\textbf{Enhancement potential for \textit{CaseMaster} usability.}}
Participants in the user study suggested areas where our system's usability could be improved. While the majority of participants appreciated the simple and clear interaction design, P5 suggested incorporating a full conversation mode. He commented, ``\textit{Sure, I can check the history, but I like ChatGPT's setup better. It lets me scroll through old conversations whenever I want, which makes it way easier to keep track of replies, even if I need to look at them again after a month.}'' Additionally, several users found the editor lacking in convenience, with half recommending the integration of an online Word feature for better usability. P6 remarked, ``\textit{I prefer the online version of Word since I can access my notes on my phone. The current editor feels a bit outdated in comparison.}'' \ouyangre{P9 further proposed, ``\textit{it would help if the logical flow of the LLM response was presented as a 'story,' maybe using mind maps or flowcharts to make it clearer.}'' Further, many participants highlighted the future value of integrating instructor involvement. P7 remarked, ``\textit{While tools like \textit{CaseMaster} are useful, in my opinion, nothing compares to the personalized guidance and interactive discussions that occur in face-to-face settings.}'' }

\section{Evaluation Study 2: Expert Evaluation}
\par After completing the controlled study, we conducted a second study to evaluate \textit{CaseMaster} from the expert perspective, involving think-aloud sessions and interviews with five experienced medical educators. Their insights highlighted opportunities to better align the tool with broader educational goals and revealed nuances that students might not be aware of, such as the appropriateness of LLM support and the overall effectiveness of the feedback mechanisms.

\subsection{Method}
\subsubsection{Participants}
\par The group of five educators (2 females, 3 males) comprises three with 2 to 3 years of teaching experience and two with 5 years of experience. Educators E1-E3 were the same individuals as participants P1, P3, and P4 from the formative study, while E4 and E5 were newly introduced to the study.

\subsubsection{Tasks}
\par
\begin{figure*}[h]
    \centering
    \includegraphics[width=\textwidth]{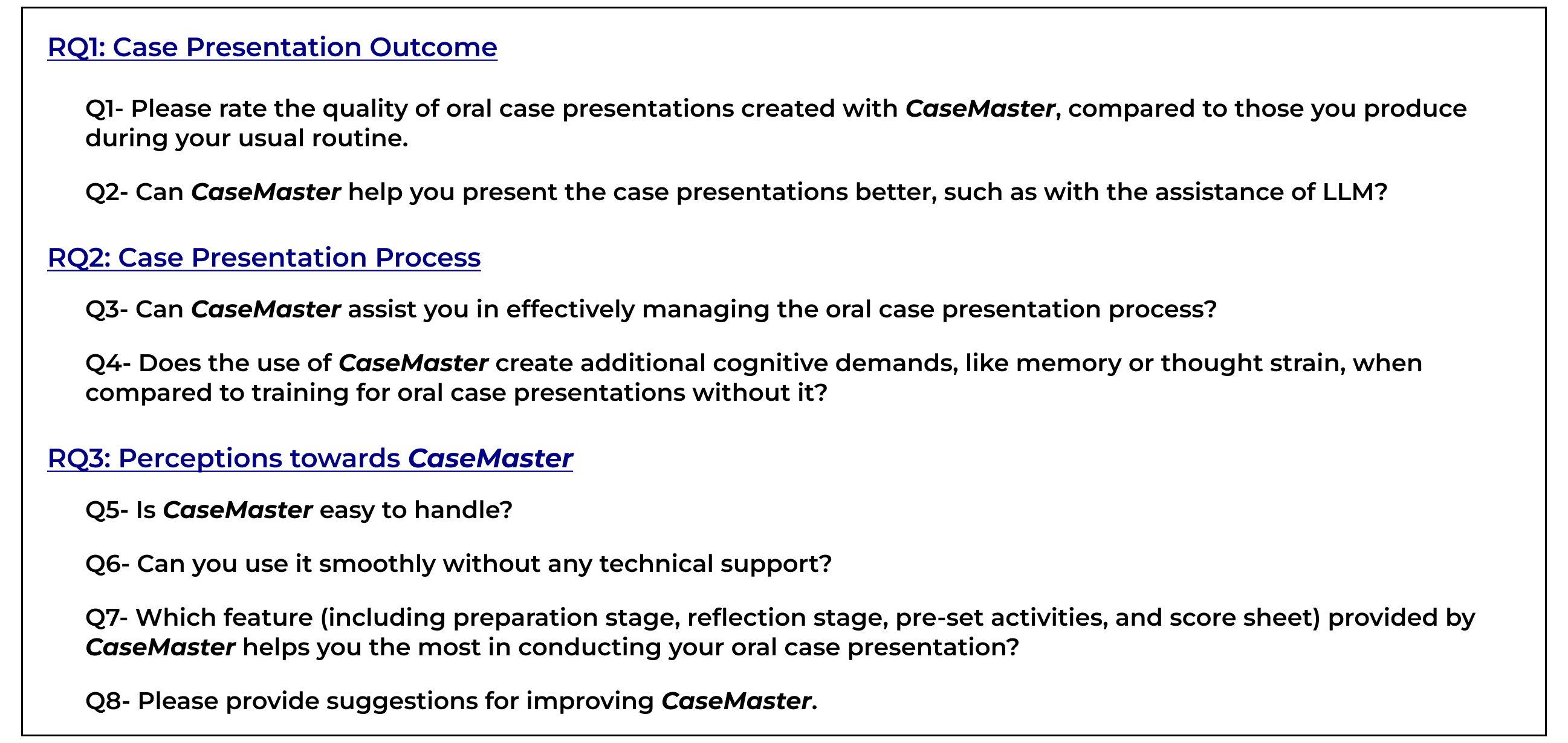}
    \caption{\ouyangre{Eight interview questions for experts, structured around the three research questions, were designed to evaluate the overall effectiveness of \textit{CaseMaster}.}}
    \label{fig:question}
\end{figure*}

 The total duration of each expert's participation was 45 minutes. Each expert's session began with an introduction to the research background and objectives. For E4 and E5, we provided an example of an OCP and SOAP format notes to facilitate a solid understanding of the concepts. This was followed by a 10-minute tutorial on using \textit{CaseMaster}. The experts were then given 20 minutes to complete tasks involving two patient cases while thinking aloud. The session concluded with a 15-minute semi-structured interview, guided by the questions outlined in \cref{fig:question}.

\subsubsection{Data collection and analysis}
\par \ouyangre{The qualitative data comprises two components: insights gathered during system use and responses from semi-structured interviews conducted afterward. All data were transcribed, and for the expert evaluation, two authors independently reviewed the transcripts to extract key insights, which were then synthesized into affinity diagrams. These diagrams were discussed to identify key themes, categorized, and cross-checked for consistency with the original transcripts. As mentioned earlier (in the data analysis of the controlled study), affinity diagrams were preferred over grounded theory due to their widespread use in HCI and interaction design~\cite{harboe2015real,lucero2015using}. After this process by the authors, the experts were individually provided with the affinity diagrams and asked to annotate and provide feedback on aspects they found either reflective of or inconsistent with their intentions and experiences. Following this, they engaged in discussions to achieve consensus and further refine the final findings.} 
\subsection{Results}
\subsubsection{Case Presentation Outcome (RQ1)}

\textbf{\ouyangre{How do experts evaluate the impact of LLM assistance on case presentations?}}
All experts acknowledged the value of the reflection stage and expressed that it could reduce their teaching workload, providing the LLM consistently delivered constructive feedback on students' solutions. E2 noted, ``\textit{... in some ways, the LLM's grading of these two patient case presentations has actually outdone some teachers, especially those from other departments.}'' However, there was some skepticism. E4 remarked, ``\textit{... I don't fully get how LLMs work, but I do know their outputs can be a bit hard to control at times, and that's something I’m worried [about]}''. Additionally, E4 mentioned a preference for intervening during presentations, stating, ``\textit{... sure, evaluating after they finish is important, but I'd rather listen to the student's [presentation] and jump in if they start going off track early on}''.

\par During the preparation stage, all experts spent minimal time using the draft assistant. In the interviews, they explained that they primarily tested the draft assistant out of curiosity, as they felt confident in their case analysis and presentation skills due to their extensive experience and familiarity with similar cases---unlike students who may be less experienced. Additionally, the case examples provided were not particularly challenging, reducing their reliance on the assistant. ``\textit{... going over a patient case has become second nature by now. As I analyze, a rough draft just starts forming in my mind automatically---it's almost effortless at this point}'' (E1).
Overall, our findings indicate that \textit{CaseMaster} significantly improved the quality of participants' case presentations. All educators valued the LLM assistance and reflection stage, noting that it streamlined their thought process and provided real-time feedback that enhanced understanding. However, certain features of \textit{CaseMaster} may not be essential in two particular scenarios: when users are already highly familiar with the patient case, or when they prefer to approach the case using a more heuristic, intuitive approach.

\subsubsection{Case Presentation Process (RQ2)}
\textbf{\ouyangre{How do experts view \textit{CaseMaster} in terms of reducing the workload?}} During interviews, experts reported minimal workload with both the baseline and \textit{CaseMaster}, attributing this to their familiarity with the case symptoms, which made the tasks relatively straightforward. They also believed that the workload for students wouldn't be overly demanding. Our main focus is identifying what contributes to students' workload during oral case presentations with \textit{CaseMaster}. E2 commented, ``\textit{... I think if students work with similar cases often, they'll get [more] comfortable managing them. And since it's not a real clinical [environment], they can practice as much as they need without feeling too much pressure}''. E4 added, ``\textit{This is an ideal training setup for students. They've got plenty of time and energy to ask the system questions and think through the process. The pressure they feel isn't the bad, stressful kind---it mostly comes from analyzing and thinking critically.}''

\subsubsection{Perceptions towards CaseMaster (RQ3)}
\textbf{\ouyangre{From the experts' perspective, are there any benefits to using \textit{CaseMaster} for OCPs?}} The experts provided positive feedback on the two-stage system design during the interviews. This design effectively enhances users' focus on preparation and reflection, thereby reducing unnecessary time spent by educators and students. E4 highlighted this benefit, stating, ``\textit{Students often need quick feedback to tweak their issues and boost [their] presentation skills, but there's usually a shortage [of resources] from educators to provide that. So in that sense, it's actually really helpful.}''

\textbf{\ouyangre{Do experts have differing opinions on the reliability of \textit{CaseMaster}?}}
Educators had mixed views on the reliability of the system's features in comparison to the baseline. The main debate focused on the preset activity, 'Provide definitions of terms.' E3 believed that the LLM-generated definitions were reasonably accurate and could be sourced, stating, ``\textit{They, [Students] have the ability to judge what's right or wrong and can check the source to see if it's reliable.}'' ``\textit{I think the LLM's content is fine as long as you can find the source. It's normal [for physicians] to have different opinions, so some variability is expected.}'' (E4) Conversely, E1 and E5 noted that LLM-provided explanations might not always match students' current understanding, potentially affecting their training. E5 explained, ``\textit{There are some concepts in medical scenarios that may be wrong or outdated for various reasons, and these could be provided as answers [by the LLMs].}'' This underscores the need for reliable dataset, a topic we will explore further in the Discussion.

\textbf{\ouyangre{Is there any enhancement potential for \textit{CaseMaster} usability?}}
During the expert interviews, experts new to the study (E4, E5) expressed a need for more detailed prompt information and the ability to customize prompts to suit their needs. E4 noted, ``\textit{LLM instruction lessons already exist in hospitals, so having the flexibility to customize the prompts would be valuable.}'' 
\ouyangre{All experts then emphasized the critical role of logic training for students throughout the process, suggesting that visualization of logic flow should be incorporated. E1 remarked, ``\textit{The current method works well, but we really need to start using visualizations to show students' logical thinking. This would not only make their thought processes clearer, but also improve the overall training experience.}''}
In line with this, experts also recognized the potential of the LLM's solution evaluation feature, which could allow educators to review evaluation and provide feedback on students' work. ``\textit{If my students had access to a tool like this for their training, I'd fully support it. I could use the reflection stage of \textit{CaseMaster} as a foundation to provide more detailed, in-depth feedback, which would greatly benefit the students.}''

\section{DISCUSSION}
\par \ouyangrevision{Our discussion first examines strategies for ensuring reliable and pedagogically grounded LLM support, then considers ways to enable flexible and adaptive use of LLMs in training, followed by the educational impact and potential generalizability of \textit{CaseMaster}, and concludes with limitations and directions for future work.}

\subsection{Designing Reliable and Pedagogically Grounded LLM Support}







\par \subsubsection{Promote Evidence-Based Critical Thinking through Structured Activities.} 
\ouyangrevision{During the preparation stage, preset activities were designed to steer students toward systematically formulating solutions.} This approach is in line with the concept of educational practices, where structured activities provide a solid foundation for developing problem-solving skills~\cite{kolb2014experiential,national2000people}. Importantly, these activities are not merely passive exercises. Unlike traditional self-study methods, which often have students passively consuming content, our approach fosters active participation by prompting students to clearly define their goals and articulate their questions effectively. This engagement is further enhanced through the integration of LLMs, which generate responses by incorporating reference materials and employing step-by-step reasoning~\cite{schimanski2024towards}.
\par \ouyangrevision{Although LLMs provide step-by-step reasoning, their responses may still contain biases or misinformation. Thus, it is critical to engage partipants in the exploration process on drafting their solution instead of solely relying on the responses provided by LLMs. In our study, students are expected to apply their knowledge to analyze, synthesize, and validate the information, ensuring that the solutions they develop are not only uniquely their own but also demonstrate critical thinking. Evidence-based critical thinking is fundamental to this process, relevant both to students’ reasoning and to evaluating the logic underlying LLM responses. While LLM responses may not always be consistent, our focus is on ensuring they have clear, reliable sources and logical reasoning, aligning with educational goals and fostering a symbiotic relationship between students and the system~\cite{Johnson2009AnEP}.}


\par \subsubsection{Advance Reflective Competencies Through Solution Comparison and LLM-Generated Feedback.} \ouyangre{Our study also emphasizes the importance of the reflection stage, where students highly value comparing different solutions, especially when the LLM highlights key issues and provides detailed feedback. However, during pre-study testing, we observed that LLMs tend to rely on affirmations to maintain user engagement, which can inadvertently obscure areas that need improvement. To address this, we implemented a structured feedback template that requires the LLM not only to mention positive aspects but also to identify specific errors or logical gaps when comparing students' solutions to references, along with detailed explanations. This approach effectively counters the tendency to overemphasize affirmations, enabling students to focus on discrepancies in logical flow and deepen their critical thinking skills. These findings echo existing research on feedback generation~\cite{Meyer2023UsingLT,Scarlatos2024ImprovingTV}. This strategy goes beyond the traditional emphasis on scores~\cite{gibbs2005conditions,biggs2022teaching}, which are often prioritized in conventional exams but fail to offer students the opportunity for meaningful engagement with their mistakes and reasoning. Instead, it promotes a more interactive, reflective process, helping students identify errors, understand their thinking, and revise their solutions.
}

\par \subsubsection{Ground LLM-Generated Content in Traditional Educational Materials and Reliable Knowledge Bases.}
One issue also highlighted in our study was that the generated content by LLM sometimes exceeded the educational level of intern medical students, presenting knowledge they had not yet learned or fully understood, reflecting the "curse of knowledge"~\cite{froyd2008faculty}. \ouyangre{This may cause confusion and frustration, hindering students' training progress and potentially undermining the overall effectiveness of the system.} 
\par \ouyangre{To address this, simplifying terminology and avoiding jargon can make the content more accessible, with key terms clearly explained to prevent confusion. One recommended method is the use of analogies to relate complex ideas to more familiar concept~\cite{chou2015using,wormeli2009metaphors}. This approach aligns with our findings, where some participants noted that drafting solutions while engaging deeply with the raw material through analogies and real-world examples was highly beneficial.} Also, our study reveals a strong preference among students for traditional educational materials—particularly those created by experienced medical educators. \ouyangrevision{This underscores the importance of integrating reliable knowledge databases, such as PubMed\footnote{\url{https://pubmed.ncbi.nlm.nih.gov/}} or ClinicalTrials\footnote{\url{https://clinicaltrials.gov/}}, into future medical education systems. These curated or verified resources could serve as reliable sources for open-source LLMs~\cite{dubey2024llama}, which can be easily customized, reducing the burden on educators to validate the generated content. Future research should focus on combining  reliable sources with adaptable models to create personalized, trainee-centered training in medical education~\cite{Bentez2024HarnessingTP, alfirevic2024custom}.
}

\subsection{Supporting Flexible and Adaptive LLM Use in Training Systems}
\par \subsubsection{Enable Flexible Customization of Prompts.} Participants valued the preset prompts provided for LLMs in the current tasks, noting their usefulness in drafting and refining content during the preparation stage and receiving constructive feedback during the reflection stage. Nevertheless, there is a growing expectation for enhanced customization of interactive features with LLMs, emphasizing the seamless integration of LLMs into workflows~\cite{yang2018investigating}.

\par Participants in the experiment, for example, preferred the ability to view and modify the built-in prompts associated with default buttons to better meet their needs. Educators also highlighted the need for mechanisms to access more detailed prompt information and customize prompts according to different student levels. For example, instead of adhering strictly to the full ``SOAP'' format, educators might wish to focus specifically on the evaluation of the plan section. One potential solution is to automate the prompting process~\cite{liu2023pre} by utilizing material metadata to provide predefined functionalities tailored to individual user requirements.

\par \subsubsection{Integrate Multi-modal Format Training.} Participants in the controlled study reported challenges in recalling the phrasing of each topic query alongside the generative responses, \ouyangre{highlighting the limitations of text-heavy outputs. Expert interviews underscored the critical need for integrating such capabilities into training systems, as current text-heavy outputs fail to support diverse learning styles and hinder practical application. }
\par \ouyangre{We held discussions on this topic, and experts first pointed out that visual tools such as binary trees, flowcharts, and concept maps are invaluable for breaking down complex ideas into more manageable components, fostering logical clarity and analytical thinking~\cite{Ghode2013InfographicsIN,musti2023virtual,praneetponkrang2014use}. These tools, when designed effectively, help students understand abstract concepts, but their effectiveness often hinges on the quality and relevance of the content. In this context, ensuring these tools align with students' cognitive styles and avoiding poorly designed content are key areas for further exploration.}
\par \ouyangre{Experts further argue that videos and simulations offer step-by-step demonstrations that enhance procedural memory. They bridge the gap between theory and practice, making them especially effective for mastering complex processes~\cite{sattar2020motivating,youssef2023learning}. However, the high costs and time-intensive production limit their widespread use, particularly in resource-constrained settings. A promising solution would be to generate videos~\cite{chen2024automatic} from existing teaching materials using LLMs, followed by strict review and validation by educators to promote accuracy and relevance. Finally, a multimodal training approach, combining visual tools, videos, and simulations, can enhance engagement, accommodate diverse styles, and foster deeper understanding~\cite{Jonassen1993StructuralKT,Mayer2019MultimediaL,oberfoell2016understanding}. Future endeavors should explore students’ needs for more finely tuned multimodal support.
}

\subsection{Broadening the Impact of \textit{CaseMaster} in Medical Education}
\ouyangrevision{Based on our findings, we highlight two key insights for medical education: first, the importance of reducing unnecessary reliance on LLMs, particularly for early-stage learners; and second, the value of integrating classical training practices alongside AI-assisted tools to maintain effective instructor-student engagement.}

\par \subsubsection{Minimize LLM Dependency in Early-Stage Medical Education.} \ouyangre{Although our participants are medical interns who have progressed beyond the foundational phases of medical education, we believe that exploring the role of LLMs in early-stage training remains valuable. In the initial phases of medical education, students are still acquiring core knowledge, clinical reasoning, and diagnostic skills. During this period, LLMs must be used cautiously, as the risk of misinformation and hallucination is heightened~\cite{da2024mitigation,huang2023survey,perkovic2024hallucinations}. If students at this stage accept inaccurate or fabricated information without critical evaluation, their developing understanding may be distorted, potentially affecting later clinical decision-making. Consequently, creating safe learning environments that mitigate these risks is essential.}
\par \ouyangre{After extensive discussions, we collectively agree that minimizing LLM dependency and fostering critical thinking skills from the outset are crucial. Our participants already have a certain level of critical thinking training from their college courses. But students in the early stages must instead be specifically trained to validate AI-generated content by cross-referencing it with trusted, authoritative sources and critically evaluating the recommendations~\cite{harrington2024mitigating}. 
\ouyangrevision{At this stage, textbooks and other trusted resources should remain the primary foundation, with LLMs serving only as supplementary tools for information retrieval rather than for clinical knowledge building or decision-making.
As students advance and gain experience, they will be better equipped to evaluate AI outputs and leverage LLMs for more advanced tasks~\cite{Farinetti2024ChatbotDU,Lyu2024EvaluatingTE}.}
In conclusion, we believe LLMs should not be relied upon in early-stage medical education and must be carefully regulated and integrated with traditional methods to mitigate potential risks.}

\par \subsubsection{\ouyangrevision{Aligning Classical Training with LLM Support to Enhance Effectiveness}} Classical training offers a foundational structure for cultivating competent medical practitioners through real-world interaction~\cite{Lee2023AdaptingHE,HuamnTapia2023CriticalTG}. \ouyangrevision{Compared to traditional methods, \textit{CaseMaster} overcomes limitations such as time constraints and inconsistent engagement by offering real-time feedback, immediate resource access, and more efficient guidance. This integration enhances the curriculum with personalized support and more structured development pathways, thereby strengthening overall training outcomes.} 


\par \ouyangrevision{Despite these benefits, open-ended responses show a persistent preference for face-to-face instruction, underscoring the distinctive value of human instructors in providing flexible and socially attuned guidance~\cite{AmreinBeardsley2007ExaminingTD}. This suggests that LLM support functions best when aligned with, rather than replacing, classical training. Our evaluation also lacked fine-grained interaction data, limiting our ability to fully understand student engagement. Future work should incorporate detailed logging within \textit{CaseMaster} to provide clearer visibility into engagement patterns and to more rigorously assess the effectiveness of LLM-supported training relative to traditional in-person instruction~\cite{Ferreira2014ANC}.
}
\subsection{\ouyangre{Generalizability of \textit{CaseMaster}: Adapting to Diverse Educational Frameworks}}
\par In the controlled study, \textit{CaseMaster} performed exceptionally well, with no participants raising concerns about the relevance of the content delivered by the LLM for students at their current educational stage. This effectiveness may be attributed to the SOAP format, a foundational theory in medical education. However, other educational frameworks, such as the OLDCART\footnote{OLDCART: Onset, Location, Duration, Character, Aggravating/Alleviating factors, Radiation, Timing} approach for assessing pain and complex symptoms and the PIE\footnote{PIE: Problem, Intervention, Evaluation} theory widely utilized in nursing, focus on different learning thresholds within various medical fields. 
This suggests that our system has the potential to be adapted across various educational levels and subjects. By collaborating with professionals from diverse departments and incorporating relevant materials to the probe, such as subject-specific grading standards, students and interns could effectively develop their skills within the framework. In addition, while our probe currently centers on OCPs, this format is merely a medium for conveying content. Expanding our focus to include training in both written and oral communication skills~\cite{Henderson2005DevelopingEP}, as well as developing a more structured format for medical written records, would be essential for enhancing these competencies.

\subsection{\ouyangre{Ethical and Privacy Considerations}}
\ouyangrevision{The deployment of LLM support in \textit{CaseMaster} introduces important ethical and privacy considerations. While the system is designed to enhance skill training, model-generated content, data practices, and interaction dynamics introduce potential risks that must be carefully managed. The key considerations include:}
\begin{itemize}
    \item \textbf{Managing Hallucinated or Misleading Content:} Although \textit{CaseMaster} does not provide diagnostic recommendations, LLM outputs may still introduce phrasing or details absent from the original case. To mitigate this risk, the system flags AI-generated segments and discloses their provenance, enabling students to identify departures from case facts. In addition, grounding model outputs in established educational materials and reliable knowledge bases further safeguards against hallucinated or misleading content~\cite{li2025accuracy,denny2023can}.
    
    \item \textbf{Fostering Calibrated Use of AI Suggestions:} Even accurate phrasing can subtly steer students toward reproducing LLM output rather than articulating their own reasoning. To sustain student autonomy, \textit{CaseMaster} prompts justification of edits and reflection on how AI suggestions align with clinical thinking. Structured activities that cultivate evidence-based critical thinking, along with solution comparison supported by targeted LLM feedback, strengthen students’ reflective competencies and encourage intentional, well-calibrated use of AI assistance~\cite{jakesch2023human,yan2024practical,wang2025genai}.
    
    \item \textbf{Safeguarding Privacy of Student Data:} The system collects audio recordings, transcripts, and draft submissions that—even when based on simulated cases—could potentially be re-identified or misused. \textit{CaseMaster} mitigates these risks through data minimization, transparent data-use communication, and plans for more granular role-based access control in future versions. These measures reflect privacy-by-design principles for educational and clinical environments~\cite{khan2024ethical}.
    
    \item \textbf{Designing for Thoughtful, Trust-Aware Interaction:} Students’ trust in AI, perceived competence, and mental models influence how they engage with LLM support. Interface strategies that foreground uncertainty, limit automation for sensitive outputs, and prompt self-reflection can steer learners toward more intentional and calibrated engagement with AI-generated content~\cite{wang2025genai}.
    
    \item \textbf{Promoting Accountability and Standards:} Developers and educators share responsibility for ensuring safe, transparent, and pedagogically aligned deployment of LLM technologies. Clear guidelines around data practices, content verification, and ongoing monitoring of learner interactions support trust and enable iterative system improvements that protect both competence development and privacy~\cite{ogunleye2024systematic,alnsour2025ai}.
\end{itemize}

\subsection{Limitation and Future work}
\par Our work has several limitations that warrant further exploration. First, the current patient data, collected by experts, focuses solely on orthopedic cases. Expanding the probe to support training cross various medical departments will require contributions from a broader range of experts or integration with online knowledge bases. To achieve this, we plan to collaborate with educators and students from diverse backgrounds to adapt and refine \textit{CaseMaster}. \ouyangrevision{Second, our study does not formally assess skill improvement through pre- and post-intervention measures or directly measure behavioral or engagement metrics, such as the number of interactions with the system. Although experts acknowledge \textit{CaseMaster} as a valuable tool for enhancing presentation quality and facilitating skill development, future work should focus on developing a train-test module specifically designed for skill-building tasks.}
Third, the sample size of 12 participants is relatively small due to time and manpower constraints. A long-term, large-scale user study would offer deeper insights into the tool's impact on sustained skill improvement over time. Fourth, the 15-minute preparation time employed in the user study, while necessary for standardization in a controlled setting, may differ from authentic OCP training scenarios that typically allow longer preparation. Moreover, while \textit{CaseMaster} primarily targets novice users, such as medical students, it also holds potential as a collaborative platform for domain experts, which could be explored in future work. Finally, to promote broader adoption, future iterations of \textit{CaseMaster} could include features that allow users to save content as digital materials for future review, or enable seamless export and sharing in slide format for presentations and collaborative purposes.
\section{CONCLUSION}
In this paper, we conducted a formative study to explore effective strategies for designing and implementing training programs that help medical students master oral case presentations (OCPs), a crucial skill in medical communication. Building on the insights from educators, we developed \textit{CaseMaster}, an interactive system probe that enhances presentation quality and reduces cognitive load for students. To assess its effectiveness, we conducted a controlled study with 12 participants, comparing \textit{CaseMaster} to a standard ChatGPT interface, alongside expert evaluation with five experienced medical educators. The results show that \textit{CaseMaster} significantly improves presentation quality and streamlines the preparation process. Our findings further reveal both limitations and opportunities, which we synthesize into design considerations to guide future development of LLM-assisted training tools in medical education.
\begin{acks}
We gratefully acknowledge the anonymous reviewers for their insightful feedback. This research was supported by the National Natural Science Foundation of China (No. 62372298), the Shanghai Engineering Research Center of Intelligent Vision and Imaging, the Shanghai Frontiers Science Center of Human-centered Artificial Intelligence (ShangHAI), and the MoE Key Laboratory of Intelligent Perception and Human-Machine Collaboration (KLIP-HuMaCo).
\end{acks}

\balance
\renewcommand{\refname}{REFERENCES}
\bibliographystyle{ACM-Reference-Format}
\bibliography{sample-base}
\newpage
\onecolumn
\appendix
\section{PROPOSED LLM-ASSISTED ACTIVITIES}
\label{sec:Pre_set}
\begin{table*}[h]
  \caption{\ouyangrevision{Proposed LLM-assisted activities for implementation in the probe (informed by design concepts from the co-design phase), including tasks such as identifying key knowledge points, reviewing medical literature, and performing logic consistency checks. Response examples demonstrate how the LLM would handle each of these tasks.}}
  \renewcommand{\arraystretch}{1.08} 
  \label{tab:activities}
\begin{tabular}{ll}
\hline
\textbf{Activity} & \textbf{Response Example} \\ \hline
Search key knowledge points &
  \begin{tabular}[c]{@{}l@{}}
  "Given the patient's symptoms, key indicators to consider include \\ 
  limited range of motion, swelling, and localized pain, which could \\ 
  suggest a potential rotator cuff injury."
  \end{tabular} \\ \hline
Review medical literature &
  \begin{tabular}[c]{@{}l@{}}
  "According to recent guidelines from the American College of \\ 
  Cardiology, initiating statin therapy in patients with LDL levels \\ 
  above 190 mg/dL is strongly recommended to reduce the risk \\ 
  of cardiovascular events."
  \end{tabular} \\ \hline
Check the logic of the content &
  \begin{tabular}[c]{@{}l@{}}
  "The treatment plan you outlined is logical, prioritizing conservative \\ 
  management such as physical therapy and NSAIDs before considering \\ 
  surgical options, which aligns with best practices for managing \\ 
  osteoarthritis of the knee."
  \end{tabular} \\ \hline
Assess the reasonableness of content &
  \begin{tabular}[c]{@{}l@{}}
  "Given the patient's active lifestyle and the severity of the ACL tear, \\ 
  the recommendation for surgical reconstruction followed by an \\ 
  aggressive rehabilitation program appears reasonable. Continued \\ 
  monitoring during recovery will be essential."
  \end{tabular} \\ \hline
Provide definitions of terms &
  \begin{tabular}[c]{@{}l@{}}
  "A comminuted fracture refers to a type of bone fracture where the \\ 
  bone is broken into multiple pieces, often requiring surgical \\ 
  intervention for proper alignment and healing."
  \end{tabular} \\ \hline
Provide specific example &
  \begin{tabular}[c]{@{}l@{}}
  "For instance, 'The patient underwent open reduction and internal \\ 
  fixation (ORIF) to stabilize the distal radius fracture following a fall.'"
  \end{tabular} \\ \hline
Explain examples in detail &
  \begin{tabular}[c]{@{}l@{}}
  "This case illustrates how early diagnosis and treatment of \\ 
  compartment syndrome, including a fasciotomy, can prevent \\ 
  long-term muscle and nerve damage in a patient with a tibial fracture."
  \end{tabular} \\ \hline
Give presentation suggestions &
  \begin{tabular}[c]{@{}l@{}}
  "When presenting this case, start with the patient's mechanism \\ 
  of injury, such as a high-energy fall, and how it led to the \\ 
  complex fracture pattern observed, to immediately capture \\ 
  the audience's attention."
  \end{tabular} \\ \hline
\end{tabular}
\end{table*}

\section{PROMPT DESIGN}
\label{sec:pro_example}
\subsection{Prompt Example for the \textit{Review Medical Literature} Activity} 
\begin{lstlisting}
{
    "role": "system",
    "content": """ 
You are a medical literature reviewer. Your task is to synthesize up-to-date, authoritative medical evidence to answer a clinical question. You should identify guideline-based recommendations and high-quality studies, extract clinically actionable findings, and produce a concise, academically written summary.

Key Considerations:
- Focus on authoritative sources such as clinical guidelines, systematic reviews, and major cohort or clinical trial evidence.
- Prioritize statements supported by strong consensus or clear clinical thresholds.
- Avoid speculation and do not infer conclusions beyond available evidence.
- Ensure the summary is concise, factual, and appropriate for a clinical or academic context.

Steps to Complete the Task:
1. Identify guideline bodies and high-quality evidence relevant to the clinical question.
2. Extract key recommendations or findings that directly address the question.
3. Evaluate the strength and clarity of the evidence.
4. Produce a short, coherent paragraph summarizing the most relevant evidence.

Output Requirements:
- Output must be a single paragraph of 3-5 sentences.
- Use formal academic tone.
- Mention guideline bodies or study types when relevant.
- Do not fabricate unsupported clinical claims.

----
Input Example:
Question: "When should lipid-lowering therapy be initiated for adults with elevated LDL cholesterol?"

Response Example:
"According to recent guidelines from the American College of Cardiology, initiating statin therapy in patients with LDL levels above 190 mg/dL is strongly recommended to reduce the risk of cardiovascular events."
"""
}
\end{lstlisting}

\subsection{Prompt Example for Rubric-Based Evaluation of Case Presentations}
\begin{lstlisting}
    {
    "role": "system",
    "content": """
You are an expert educational evaluator specializing in precise and objective grading of student oral case presentations. Your task is to evaluate a student's solution against a reference answer using a predefined scoring rubric. Each section is scored on a scale of 0 to 3, with higher scores indicating better performance. Some sections also require Yes/No evaluations. Avoid speculating beyond what the student wrote.

Key Considerations:
- Evaluate the student's solution against the reference answer.
- Evaluate each rubric dimension using evidence present in the student solution and the reference answer.
- Provide concise, objective justifications citing specific content from the student solution.
- Highlight both strengths and areas needing improvement.
- Avoid assumptions beyond the content of the student solution and reference answer.
- Maintain clarity and professional tone suitable for clinical education contexts.

Scoring Criteria:
  - History:
    - Timing and characterization of symptoms: 0 to 3
    - Includes pertinent facts but excludes extraneous information: 0 to 3
  - Important Information:
    - Relevance and focus of medical history, family history, surgical history, allergies, medications, and social history: 0 to 3
    - Avoids unnecessary details / separate review of systems: Yes/No
  - Physical Examination:
    - Prioritization of vital signs: Yes/No
    - Focused examination relevant to diagnosis: 0 to 3
  - Labs:
    - Inclusion of essential laboratory data: 0 to 3
    - Understanding of relevant labs: 0 to 3
  - Assessment and Plan:
    - Clarity of the synthesis statement: 0 to 3
    - Logic and sufficiency of differential diagnoses: 0 to 3
  - General and Style:
    - Duration (5-8 minutes): 0 to 3
    - Organization in logical order: 0 to 3
    - Use of distractors (uhs, uhms, ahs): 0 to 3
    - Strength of argument / Makes a case: 0 to 3


Steps to Complete the Task :
1. Review the student's presentation carefully.
2. Compare the student's content against the reference answer.
3. For each rubric dimension, identify supporting or missing evidence.
4. Assign a numerical rating or Yes/No according to the rubric.
5. Provide a brief justification (1-2 sentences) citing specific content from the student's response and referencing the standard answer as appropriate.
6. Output your evaluation as a structured JSON object.

Output Format:
- JSON object with keys as rubric dimensions.
- Each key maps to an object containing:
    - 'score': numerical rating or Yes/No
    - 'justification': brief text citing student evidence

Input Example:
Student Solution: "The patient has a history of hypertension and presents with elevated blood pressure. Lifestyle modifications were recommended, and a low-dose ACE inhibitor was prescribed...."

Rubric Dimensions: [
  "Timing and characterization of symptoms",
  "Includes pertinent facts but excludes extraneous information needed to establish and modify a differential",
  "Relevance and focused reporting of medical history, family history, surgical history, allergies, medications, and social history",
  "Avoids a separate review of system",
  "Vital signs first",
  "Focused physical examination relevant to the diagnosis, includes data necessary for the differential diagnosis but excludes extraneous information",
  "Includes laboratory data essential to the diagnosis and excludes irrelevant data",
  "Demonstrates understanding of which labs are relevant and which are not",
  "Synthesis statement",
  "Assessment includes a list of at least three differential diagnoses with arguments for and against each. Arguments are succinct",
  "Duration",
  "Organization in the proper order",
  "Use of distractors (uhs, uhms, ahs)",
  "Makes a case"
]

Response Example (JSON):
{
  "Timing and characterization of symptoms": {
    "score": 2,
    "justification": "The student effectively described the timing and characterization of symptoms, including the initial trauma and subsequent exacerbation of pain and swelling."
  },
  "Includes pertinent facts but excludes extraneous information needed to establish and modify a differential": {
    "score": 1,
    "justification": "The presentation included pertinent facts regarding Lee's history and management but could benefit from excluding some extraneous details that do not directly impact the differential diagnosis."
  },
  "Relevance and focused reporting of medical history, family history, surgical history, allergies, medications, and social history": {
    "score": 3,
    "justification": "The student provided a comprehensive and focused report of the patient's medical history, demonstrating a clear understanding of the relevant aspects."
  },
  "Avoids a separate review of system": {
    "score": "Yes",
    "justification": "The student successfully avoided a separate review of systems, maintaining a concise and relevant presentation."
  },
  "Vital signs first": {
    "score": "No",
    "justification": "The student did not prioritize vital signs at the beginning of the presentation, missing an important aspect of the clinical assessment."
  },
  "Focused physical examination relevant to the diagnosis, includes data necessary for the differential diagnosis but excludes extraneous information": {
    "score": 2,
    "justification": "The student conducted a focused physical examination related to the diagnosis, incorporating essential data for the differential diagnosis while maintaining relevance."
  },
  "Includes laboratory data essential to the diagnosis and excludes irrelevant data": {
    "score": 0,
    "justification": "The presentation did not include essential lab data in the solution, missing crucial information for the diagnosis."
  },
  "Demonstrates understanding of which labs are relevant and which are not": {
    "score": 1,
    "justification": "The student showed some understanding of relevant labs but could improve by clearly differentiating between essential and non-essential data."
  },
  "Synthesis statement": {
    "score": 3,
    "justification": "The student provided a clear and concise synthesis statement, effectively summarizing the key elements of the case presentation."
  },
  "Assessment includes a list of at least three differential diagnoses with arguments for and against each. Arguments are succinct": {
    "score": 2,
    "justification": "The student presented a reasonable assessment with multiple differentials, although some arguments could be further strengthened and more succinctly articulated."
  },
  "Duration": {
    "score": 3,
    "justification": "The student's presentation duration fell within the appropriate range, allowing for a comprehensive but concise evaluation."
  },
  "Organization in the proper order": {
    "score": 1,
    "justification": "The organization could be improved by aligning the flow of information more closely with the logical progression of a case presentation."
  },
  "Use of distractors (uhs, uhms, ahs)": {
    "score": 3,
    "justification": "The student effectively maintained a fluent presentation without unnecessary distractors, contributing to the clarity of the communication."
  },
  "Makes a case": {
    "score": 3,
    "justification": "The student successfully built a strong case, integrating clinical findings with diagnostic data to support the proposed management plan."
  }
}
"""
}

\end{lstlisting}

\newpage
\section{QUESTION STATEMENTS ASSESSING TRUST AND CONCERNS ABOUT THE SYSTEM (JIAN’S TRUST SCALE)}
\label{sec:system_trust}

\begin{table*}[h]
    \centering
    \resizebox{\linewidth}{!}{%
    \begin{tabular}{p{4cm}p{12cm}}
        \hline
        \textbf{Aspect} & \textbf{Question Item} \\
        \hline
        System Vigilance & \textit{I’m concerned that the system might not be vigilant enough to handle unexpected problems.} \\
        Potential Negative Effects & \textit{I am concerned that the system could have negative effects on OCP training.} \\
        Ethical Standards & \textit{I feel confident that the system will respect user privacy and uphold ethical standards in its operation.} \\
        Output Reliability & \textit{I trust the system to deliver reliable and consistent results.} \\
        User Commitment & \textit{I trust the system enough to invest time and resources in using it for the long term, despite occasional errors.} \\
        \hline
    \end{tabular}}
    \label{tab:system_trust}
\end{table*}

\section{SYSTEM WALKTHROUGH}
\label{sec:walkthrough}
\newpage

\begin{figure}[H]
    \centering
    \includegraphics[width=\textwidth]{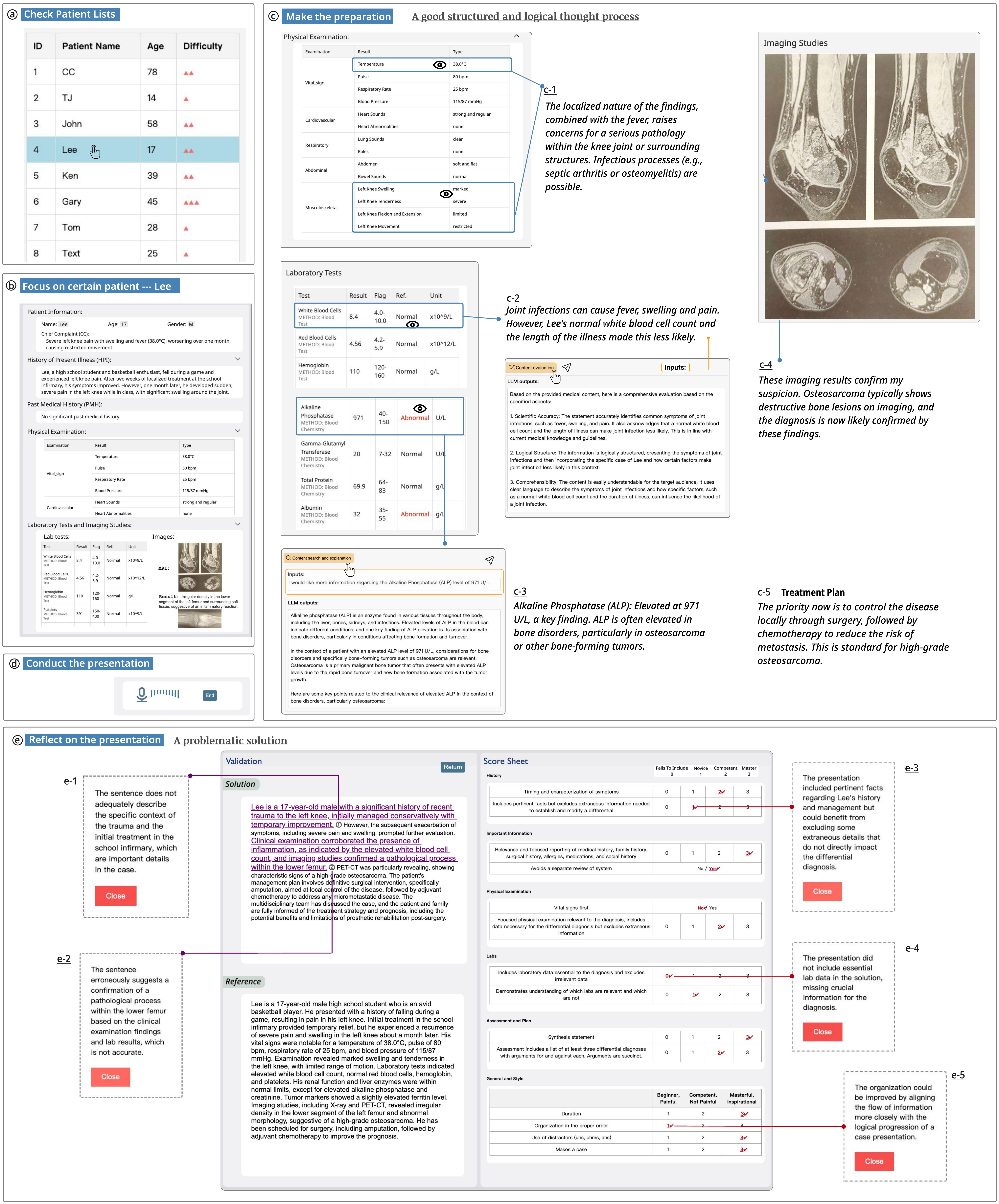}
    \vspace{-7mm}
    \caption{The process of conducting case presentation training with \textit{CaseMaster} includes the following steps: (a) Reviewing the patient list, (b) Focusing on a specific patient, ``Lee'', (c) Demonstrating a well-structured and logical thought process through preset activities and organized preparation, (d) Conducting the oral case presentation, and (e) Highlighting a problematic solution, where another user gets feedback on the presentation.}
    \label{fig:walkfigure}
\end{figure}

\twocolumn
\par We demonstrate the use of \textit{CaseMaster} through a detailed workflow scenario, as illustrated in \cref{fig:walkfigure}. The scenario involves two users, Mary and Bob: one who performs comprehensive preparation and another who engages in reflective analysis. They start by reviewing the patient list (\cref{fig:walkfigure} (a)) and select a specific patient, ``Lee'' (\cref{fig:walkfigure} (b)). Lee is a 17-year-old male experiencing severe left knee pain, swelling, and fever (38.0°C). His symptoms have progressively worsened over the past month, significantly restricting his movement. Lee, a high school student and basketball player, initially injured his knee during a game. After receiving localized treatment at the school infirmary for two weeks, his symptoms improved. However, one month later, he experienced a sudden onset of severe pain and swelling in the knee while in class. 
\par Mary began her preparation (\cref{fig:walkfigure} (c)) with a Physical Examination. She observed that the patient had a fever of 38.0°C, although other vital signs were relatively normal. The left knee was swollen, exhibited severe tenderness, and had limited movement, indicating possible joint or bone involvement. She documented these findings and noted that,
\begin{quote}
    ``\textit{The localized nature of the findings, combined with the fever, raised concerns for a serious pathology within the knee joint or surrounding structures. Infectious processes, such as septic arthritis or osteomyelitis, were possible (\cref{fig:walkfigure} (c-1)).}''
\end{quote}

\par Next, she checked the Laboratory Tests. The patient's White Blood Cells (WBC) count was normal at \(8.4 \times 10^9/\text{L}\). She observed,
\begin{quote}
    ``\textit{Joint infections can cause fever, swelling, and pain. However, Lee’s relatively normal WBC count and the longer history reduced the likelihood of this (\cref{fig:walkfigure} (c-2)).}''
\end{quote}

\par She then proceeded to query the draft assistant, using the ``Content Evaluation'' preset activity. The assistant verified that her reasoning was both logical and scientifically sound.

\par Upon reviewing the Alkaline Phosphatase (ALP) test results, she observed an elevated level of \(971 \, \text{U/L}\), which was abnormal. This critical finding redirected her diagnostic focus. Recognizing that elevated ALP is commonly associated with bone-related disorders, particularly bone tumors like osteosarcoma, she sought additional information about the ALP test. Using the ``Content Search and Explanation'' preset activity, she received a comprehensive explanation from the draft assistant. She concluded that,
\begin{quote}
    ``\textit{Alkaline Phosphatase (ALP) is elevated at \(971 \, \text{U/L}\), a key finding. ALP is often elevated in bone disorders, particularly in osteosarcoma or other bone-forming tumors (\cref{fig:walkfigure} (c-3)).}''
\end{quote}
\par She then reviewed the imaging studies, which revealed irregular density and abnormal morphology in the lower femur. She noted that these findings were consistent with potential bone pathology, reinforcing her concern for a possible bone tumor. She stated,
\begin{quote}
    ``\textit{These imaging results confirmed my suspicion. Osteosarcoma typically shows destructive bone lesions on imaging, and the diagnosis was now likely confirmed by these findings (\cref{fig:walkfigure} (c-4)).}''
\end{quote}
\par Finally, she summarized her analysis and developed a treatment plan, concluding that:
\begin{quote}
    ``\textit{The priority now was to control the disease locally through surgery, followed by chemotherapy to reduce the risk of metastasis. This is the standard approach for high-grade osteosarcoma (\cref{fig:walkfigure} (c-5)).}''
\end{quote}

\par Bob reviewed his presentation and identified several issues (\cref{fig:walkfigure} (e)). Initially, he concentrated on the \textit{Validation} section, addressing each poorly constructed sentence individually.
\begin{itemize}
    \item \cref{fig:walkfigure} (e-1) highlights that, 
    \begin{quote}
        ``\textit{The sentence does not adequately describe the specific context of the trauma and the initial treatment in the school infirmary, which are important details in the case.}''
    \end{quote}
    \item \cref{fig:walkfigure} (e-2) shows, 
    \begin{quote}
        ``\textit{The sentence erroneously suggests confirmation of a pathological process within the lower femur based on clinical examination findings and lab results, which is not accurate.}''
    \end{quote}
\end{itemize}
\par Bob then examined the reference answer to gain a deeper insight into effective presentation skills. He proceeded to review the \textit{Score Sheet}:
\begin{itemize}
    \item \cref{fig:walkfigure} (e-3) notes, 
    \begin{quote}
        ``\textit{The presentation included pertinent facts regarding Lee's history and management but could benefit from excluding some extraneous details that do not directly impact the differential diagnosis.}''
    \end{quote}
    \item \cref{fig:walkfigure} (e-4) points out that, 
    \begin{quote}
        ``\textit{The presentation did not include essential lab data in the solution, missing crucial information for the diagnosis.}''
    \end{quote}
    \item \cref{fig:walkfigure} (e-5) observes that, 
    \begin{quote}
        ``\textit{The organization could be improved by aligning the flow of information more closely with the logical progression of a case presentation.}''
    \end{quote}
\end{itemize}
\par Bob felt he had gained substantial knowledge and was confident in his understanding. He looked forward eagerly to the next study.











\end{document}